\documentclass[aps,prd,showpacs,nofootinbib,superscriptaddress,twocolumn,amsmath]{revtex4-1}
\usepackage{graphicx,amssymb,braket}
\usepackage{soul,bbm,dcolumn,booktabs}
\usepackage{multirow,array,float,enumitem}
\usepackage{lmodern,dsfont,microtype,nicefrac,microtype}
\usepackage{rotating,adjustbox,bm,xparse}
\usepackage[dvipsnames]{xcolor}
\usepackage[utf8]{inputenc}
\usepackage{upgreek}
\usepackage{array}
\usepackage{booktabs}
\usepackage{enumitem}
\usepackage{pifont}
\usepackage{stackengine}

\usepackage{sidecap}
\newcolumntype{C}{>{$}c<{$}}
\AtBeginDocument{
\heavyrulewidth=.08em
\lightrulewidth=.05em
\cmidrulewidth=.03em
\belowrulesep=.65ex
\belowbottomsep=0pt
\aboverulesep=.4ex
\abovetopsep=0pt
\cmidrulesep=\doublerulesep
\cmidrulekern=.5em
\defaultaddspace=.5em
}
\newcommand{\be}{\begin{eqnarray}}
\newcommand{\ee}{\end{eqnarray}}

\def\fm {\mathop{\hbox{fm}}}
\def\MeV {\mathop{\hbox{MeV}}}
\def\GeV {\mathop{\hbox{GeV}}}

\def\beq{\begin{equation}}
\def\eeq{\end{equation}}
\def\beqs#1\eeqs{\beq\begin{split} #1 \end{split}\eeq}

\def\comment#1{}

\def\ket#1{\left| #1 \right\rangle}
\def\bra#1{\left\langle #1 \right|}

\newcolumntype{L}{>{$}l<{$}} 
\newcolumntype{S}{>{\footnotesize $}l<{$\normalsize}} 

\def\*#1{\mathbf{#1}}

\newcolumntype{R}[2]{%
    >{\adjustbox{angle=#1,lap=\width-(#2)}\bgroup}%
    c%
    <{\egroup}%
}

\newcommand{\lvec}{\bm{l}}
\newcommand{\tlvec}{\tilde{\bm{l}}}
\newcommand{\pvec}{\bm{p}}
\newcommand{\tpvec}{\tilde{\bm{p}}}

\newcommand{\kvec}{\bm{k}}
\newcommand{\tkvec}{\tilde{\bm{k}}}

\newcommand{\tPvec}{\tilde{\bm{P}}}

\newcommand{\qvec}{\bm{q}}
\newcommand{\tqvec}{\tilde{\bm{q}}}

\def\*#1{\bm{#1}}
\DeclareMathOperator{\diag}{diag}

\setitemize[1]{leftmargin=12pt}

\usepackage[colorlinks=true,backref=false,linktocpage=true,
pdfpagemode=UseOutlines]{hyperref}
\hypersetup{%
              bookmarksnumbered=true,
              linkcolor=NavyBlue,
              citecolor=NavyBlue,
              urlcolor=NavyBlue,
              pdftitle = {},
              pdfsubject = {},
              pdfauthor = {},
              pdfkeywords = {}
}

\begin{document}

\title{
Three-body interactions from the finite-volume QCD spectrum}

\author{Ruair\'{i}~Brett}
\email{rbrett@gwu.edu}
\affiliation{The George Washington University, Washington, DC 20052, USA}
\author{Chris~Culver}
\email{C.Culver@liverpool.ac.uk}
\affiliation{Department of Mathematical Sciences, University of Liverpool, Liverpool L69 7ZL, United Kingdom}
\author{Maxim~Mai}
\email{maximmai@gwu.edu}
\affiliation{The George Washington University, Washington, DC 20052, USA}
\author{Andrei~Alexandru}
\email{aalexan@gwu.edu}
\affiliation{The George Washington University, Washington, DC 20052, USA}
\affiliation{Department of Physics, University of Maryland, College Park, MD 20742, USA}
\author{Michael~D\"{o}ring}
\email{doring@gwu.edu}
\affiliation{The George Washington University, Washington, DC 20052, USA}
%
\author{Frank~X.~Lee}
\email{fxlee@gwu.edu}
\affiliation{The George Washington University, Washington, DC 20052, USA}
%

\begin{abstract}
We perform a fit of the finite-volume QCD spectrum of three pions at maximal isospin to constrain the three-body force. We use the unitarity-based relativistic three-particle quantization condition, with the GWUQCD spectrum obtained at 315~MeV and 220~MeV pion mass in two-flavor QCD.
For the heavier pion mass we find that the data is consistent with a constant contact term close to zero, whereas for the lighter mass we see a statistically significant energy dependence in tension with the prediction of leading order ChPT.
Our results also suggest that with enough three-body energy levels, the two-body amplitude could be constrained.
\end{abstract}

\pacs
{
12.38.Gc, 
14.40.-n, 
13.75.Lb  
}

\maketitle

\setlength{\parskip}{8pt}

\section{Introduction}\label{sec:intro}

It is a long-term quest of nuclear physics to understand hadron interactions as they emerge from quark and gluon dynamics. The main challenge lies in the fact that perturbation theory fails at low energies, because the interactions are strong. Lattice QCD~(LQCD) offers a non-perturbative method which has quarks and gluons as fundamental degrees of freedom while keeping all systematics under control. LQCD calculations are performed in a finite volume and in Euclidean time, leaving only indirect methods to study real-time infinite-volume scattering.  The relation between finite-volume spectrum and infinite-volume scattering amplitudes is called quantization condition, which has been known for two-hadron systems since the pioneering work of L\"uscher~\cite{Luscher:1990ux}.  The last decade has witnessed significant progress using this approach for a variety of interacting two-particle systems. Only recently has the quantization condition been extended to the three-hadron sector.

Recent theoretical advances~\cite{%
Kreuzer:2008bi,
Kreuzer:2009jp,
Bour:2011ef,
Kreuzer:2012sr,
Polejaeva:2012ut,
Briceno:2012rv,
Meissner:2014dea,
Hansen:2014eka,
Hansen:2015zga,
Agadjanov:2016mao,
Guo:2016fgl,
Hansen:2016fzj,
Briceno:2017tce,
Meng:2017jgx,
Mai:2017bge,
Guo:2017ism,
Hammer:2017uqm,
Hammer:2017kms,
Guo:2018ibd,
Doring:2018xxx,
Guo:2018xbv,
Romero-Lopez:2018rcb,
Briceno:2018aml,
Guo:2019hih,
Romero-Lopez:2019qrt,
Zhu:2019dho,
Romero-Lopez:2020rdq,
Guo:2020spn,
Guo:2020wbl,
Guo:2019ogp,
Blanton:2020gmf,
Blanton:2020gha, 
Blanton:2020jnm
} 
of the three-body formalism as well as related numerical studies~\cite{%
Kreuzer:2010ti,
Roca:2012rx,
Bour:2012hn,
Jansen:2015lha,
Guo:2017crd,
Mai:2018djl,
Klos:2018sen,
Briceno:2018mlh,
Blanton:2019vdk,
Mai:2019fba,
Hansen:2020otl,
Muller:2020vtt,
Guo:2020kph,
Guo:2020ikh,
Pang:2020pkl} 
(see Refs.~\cite{Hansen:2019nir,Rusetsky:2019gyk} for reviews) open a possibility for studying hadronic processes that involve three-body interactions from first principles. Many scattering channels of interest receive contributions from three-particle states. For example in the meson sector the $a_1(1260)$ resonance, seen experimentally in $\tau$ decays, couples first to $\rho\pi$ and $\sigma\pi$, and then due to the instability of the $\rho$ and $\sigma$ mesons to a final state with three pions. In the baryonic sector an example where three-body states are relevant is the Roper resonance $N(1440)1/2^+$ which has both two- and three-particle final states, as it decays to $\pi N$ and $\pi\pi N$. 

So far most of the effort in lattice QCD calculations of three-hadron state energies has been concentrated on pure three-meson systems with maximal isospin (three pions or three kaons) where relatively precise finite-volume spectra have been calculated~\cite{Detmold:2008fn, Detmold:2008yn, Horz:2019rrn,Culver:2019vvu,Fischer:2020jzp,Hansen:2020otl, Alexandru:2020xqf} and the formalism connecting it to infinite-volume amplitudes is better understood~\cite{%
Hansen:2014eka,
Mai:2017bge,
Hammer:2017uqm,
Hammer:2017kms,
Mai:2018djl,
Doring:2018xxx,
Briceno:2018mlh,
Mai:2019fba,
Blanton:2019vdk,
Hansen:2020otl,
Muller:2020vtt,
Guo:2020kph,
Muller:2020wjo
}. 
In the pioneering work by the NPLQCD collaboration in the $3\pi$ and $3K$ sectors~\cite{Detmold:2008fn, Detmold:2008yn} threshold energy eigenvalues at different pion masses were determined and a threshold expansion~\cite{Detmold:2008gh, Beane:2007qr, Tan:2007bg} was performed allowing for the first determination of a three-body force. 

Finite-volume formalisms for excited states were developed later. In Ref.~\cite{Mai:2021lwb} different approaches are reviewed in more detail, including a timeline and a comparison of the methods. In the following we provide a very brief overview. Among other recent developments~\cite{Guo:2020kph, Romero-Lopez:2020rdq}, there are two relativistic formalisms available for three-particle scattering usually referred to as relativistic finite-volume unitarity~(FVU)~\cite{Mai:2017bge} and relativistic effective field theory~(RFT)~\cite{Hansen:2014eka} approaches.  The former uses unitarity of the $S$-matrix as a guiding principle, while the latter relies on the re-summation of diagrams. A key ingredient to these formalisms is the parameterization of two and three-body scattering amplitudes. 
The two-body input can be determined from experiment in combination with chiral extrapolations, but lattice QCD data, together with the quantization conditions, can also be used. 

The first determination of the three-body force from the NPLQCD data~\cite{Detmold:2008fn} using such a formalism was achieved in Ref.~\cite{Mai:2018djl} with the FVU framework. There, within the uncertainties of the lattice data, the three-body force was found to be zero. This study contains also the first predictions for excited levels at different pion masses. Later, in Ref.~\cite{Horz:2019rrn} excited levels for different boosts and irreducible representations (irreps) were calculated that led to a refined determination of the three-body force with the RFT formalism~\cite{Blanton:2019vdk}. In particular, the three-body force was found to be non-zero, and even attempts to determine its energy dependence could be made. Within the FVU formalism~\cite{Mai:2019fba} the data of Ref.~\cite{Horz:2019rrn} were predicted using only chiral extrapolations of two-body input up to next-to-leading order (NLO) and assuming a vanishing three-body force.

The GWUQCD collaboration calculated the three-$\pi^+$ spectrum for different quark masses, box geometries, and boosts, mapping out a plethora of states and comparing also to FVU predictions that were made under the assumption of vanishing three-body forces~\cite{Culver:2019vvu}. The agreement found was fair.
Subsequently, the ETMC collaboration calculated the three-$\pi^+$ spectrum at three different pion masses, including the physical point for the first time~\cite{Fischer:2020jzp}. The extraction of the three-body force with the RFT formalism was compared with the leading order~(LO) Chiral Perturbation Theory~(ChPT) prediction for the three-to-three process. Similar to Ref.~\cite{Blanton:2019vdk}, the three-body term was found to be non-zero, its energy-independent part being in qualitative agreement with LO ChPT, but its energy-dependent part not.

The evaluation of the infinite-volume three-body amplitude is a challenge by itself. For example, the FVU framework was recently extended to the infinite volume (albeit without lattice input), in order to study the decay {$a_1(1260)\to \pi\rho$} in coupled S- and D-waves~\cite{Sadasivan:2020syi}.
For three positive pions, the infinite-volume scattering amplitude was solved in Ref.~\cite{Hansen:2020otl} in the RFT framework, for the first time with actual lattice input. The input was calculated by the Hadron Spectrum collaboration and the analysis included an extraction of the three-body force.
Within uncertainties and different fit strategies/parametrizations tried, the three-body term ${\cal K}_\text{3,iso}$ was found to be compatible with zero.

Obviously, the role of three-body forces is not settled. In this paper, we continue the investigation of the three-body force with the FVU formalism. In particular, the mentioned discrepancy of FVU predictions with GWUQCD data ($\chi^2_\text{dof}\approx 2.68$)~\cite{Culver:2019vvu} could be the result of a finite three-body term.
Therefore, we perform here a fit to the lattice data of Ref.~\cite{Culver:2019vvu} using the FVU formalism with special focus on the three-body contact term. In Section~\ref{sec:lattice_data} we review the extraction of the three-pion finite-volume spectrum from the GWUQCD collaboration.  Then in Section~\ref{sec:qc}, we review the three-body quantization condition. We include technical implementation details such as parameterizations of the two- and three-body scattering amplitudes, and establish a connection to a three-to-three contact term.  In Section~\ref{sec:results} we present the results of various fit scenarios. Finally in Section~\ref{sec:conclusions} we discuss the impact of the results.

\section{Finite-volume QCD spectrum}
\label{sec:lattice_data}

\begin{table}[t]   
\begin{tabular}{@{}*{13}{>{$}l<{$}}@{}}                                                     
\toprule                                                                                               
\text{Label}~& N_t\times N_{x,y}^2\times N_z & ~\eta~ & ~a[\fm]~   & ~N_\text{cfg}~  & ~am_{\pi}~ \\
\midrule                                                                                               
\mathcal{E}_1&48\times24^2\times24  &  1.00  & 0.1210(2)(24) & 300   & 0.1931(4)\\ 
\mathcal{E}_2&48\times24^2\times30  &  1.25 & -        & -     &    0.1944(3)\\
\mathcal{E}_3&48\times24^2\times48  &  2.00  & -        & -     &    0.1932(3)\\
\mathcal{E}_4&64\times24^2\times24  &  1.00  & 0.1215(3)(24)& 400  & 0.1378(6) \\
\mathcal{E}_5&64\times24^2\times28  &  1.17 &      -   & -     & 0.1374(5)\\
\mathcal{E}_6&64\times24^2\times32 &  1.33 &      -   & -     & 0.1380(5)\\
\bottomrule                                                                                            
\end{tabular}
\caption{
\label{table:gwu_lattice}
Details of the $N_f=2$ ensembles used in this study.  Here $\eta$ is elongation in the $z$-direction, $a$ the lattice spacing,  $N_{\text{cfg}}$ the number of Monte Carlo configurations.  Ensembles $\mathcal{E}_1, \mathcal{E}_2, \mathcal{E}_3$ have a pion mass $m_\pi \approx 315~\MeV$, while $\mathcal{E}_4, \mathcal{E}_5, \mathcal{E}_6$ have a pion mass $m_\pi \approx 220 \MeV$.}    
\end{table} 

Here we review briefly the details of the calculation of the finite-volume spectrum performed by our collaboration, referring the reader to Ref.~\cite{Culver:2019vvu} for additional material.  The ensembles employed use $N_f=2$ dynamical fermions, with two sets of quark masses tuned so that the pions have masses of $315\MeV$ or $220\MeV$. These data sets were used to compute the two-pion spectrum in all three isospin channels~\cite{Pelissier:2012pi,Guo:2016zos,Guo:2018zss,Culver:2019qtx}, the three-pion $I=3$ channel~\cite{Culver:2019vvu} and the $K^-K^-K^-$ channel~\cite{Alexandru:2020xqf}. The quark propagators are estimated using LapH smearing~\cite{Peardon:2009gh} and an optimized code is employed to compute the required matrix inversions~\cite{Alexandru:2011ee}. 
The parameters describing the ensembles in this study are listed in Table~\ref{table:gwu_lattice}. The lattice spacing was tuned using the Wilson flow parameter $t_0$. For details about tuning the lattice spacing and other parameters we refer the reader to Ref.~\cite{Niyazi:2020erg}.

For each pion mass, we calculated the $I=3$ three-pion finite-volume spectrum on three different ensembles, which feature two different geometries.  There is one cubic volume and two volumes which are elongated in the $z$-direction, which is also the direction of the boost when calculating the energies of moving states. The advantage of using elongated ensembles is an increase in energy resolution, since the momenta are quantized in units of $2\pi/L$. For ensemble $\mathcal{E}_3$, we are able to extract $16$ energy levels below the inelastic threshold, compared to $3$ in the cubic volume at the same pion mass. The numerical cost of generating these ensembles is reduced since the volume increases linearly with the elongation, as opposed to a cubic increase for symmetric boxes.
In total we have 30 energy levels below $5m_\pi$ across the six ensembles listed in Table~\ref{table:gwu_lattice}. They form the basis for our quantitative analysis of the three-body interaction in this work. Precise values for the levels can be found in Appendix B of Ref.~\cite{Culver:2019vvu}.

Regardless of geometry, the finite volume introduces a critical change to the angular momentum quantum number of the system of interest.  In the infinite volume, the symmetry group of rotations is $SO(3)$, and the quantum number of the states are labelled by the angular momentum $l$, the irreducible representations~(irreps) of the group.  For a finite volume the symmetry group is reduced from $SO(3)$ to $O_h$ in the case of cubic volumes, and $D_{4h}$ in the case of elongated volumes.  If the system is at non-zero total momentum the relevant symmetry group is $C_{4v}$, provided the boost is along the direction of any spatial elongation.  In this study we only consider a boost of $P=[001]$, which is aligned with the elongation.  The rotation quantum number of states in a finite volume are thus labelled by the irreps of the respective symmetry groups.  Here we focus only on states that overlap with $l=0$, which will be labelled by $A_{1u}$ for symmetry groups $O_h$ and $D_{4h}$, or labelled by $A_2$ for the $C_{4v}$ symmetry group. 
For a full discussion on the relation between angular momentum $l$ and irreps of the finite-volume symmetry groups in elongated boxes we refer the reader to Ref.~\cite{Lee:2017igf}.

\section{Quantization Condition}\label{sec:qc}

The dynamics of a (multi-)hadron system is accessed in lattice QCD by calculating correlation functions on a discretized Euclidean space-time in a finite volume. The so-called discretization effects are related to the finite lattice spacing $a$. In principle, a continuum limit $a\to0$ needs to be taken to relate the lattice QCD results to the physical (real-world) quantities. Since only a finite number of lattice sites can be considered in any practical calculation, the system of interest is necessarily evaluated in a finite volume. Imposing boundary conditions restricts the allowed momenta in this system. For example, in a cubic volume of side length $L$, the frequently applied (also here) periodic boundary conditions yield an infinitely large discrete set of allowed momenta $\mathcal{S}_L=\{(2\pi/L)\bm{n}|\bm{n}\in \mathds{Z}^3\}$. Unavoidably, using three-momenta from such a set changes the infinite-volume spectrum fundamentally, making it a discrete set of energy eigenvalues. This holds for any finite $L$, such that a simple extrapolation ($\lim_{L\to \infty}$) is not useful, calling for a non-trivial mapping between infinite- and finite-volume. Such mappings are referred to as \textit{quantization conditions}. An alternative possibility to use ordered double limit~\cite{DeWitt:1956be}
techniques to extract (complex-valued) amplitudes was explored in Ref.~\cite{Agadjanov:2016mao}, with related techniques in Refs.~\cite{Hansen:2017mnd,Guo:2020ikh,Briceno:2020rar,Bulava:2019kbi}. The goal of this section is to reiterate the form of the relativistic three-body quantization condition derived in Refs.~\cite{Mai:2017bge,Mai:2018djl,Mai:2019fba,Culver:2019vvu}, unifying and simplifying the nomenclature. 

As one of the currently available relativistic formulations of the three-body quantization condition (see Refs.~\cite{Hansen:2019nir,Rusetsky:2019gyk} for reviews) the FVU approach derives from the relativistic unitary three-body formalism~\cite{Mai:2017vot} in infinite volume. The formalism differs from the diagrammatic approach followed in Ref.~\cite{Hansen:2014eka}, but yields an equivalent finite-volume spectrum given the same input, as shown using time ordered perturbation theory in Ref.~\cite{Blanton:2020jnm}. Also in the infinite volume both formalisms yield a unitary three-body amplitude~\cite{Jackura:2019bmu,Briceno:2019muc}. Both formulations are currently being applied to a variety of calculations of simpler three-hadron systems such as $3\pi^+$~\cite{Mai:2018djl,Blanton:2019vdk,Mai:2019fba,Culver:2019vvu,Blanton:2019igq,Hansen:2020otl} or $3K^-$ \cite{Alexandru:2020xqf}. However, the practical implementation of each differs.
See also Ref.~\cite{Guo:2020kph} for a calculation based on combination of a variational approach and the Faddeev method, including relativistic effects for the pions and lattice spacing effects.
In the following, we will demonstrate and discuss an example of contrasting representations of the three-body contact term.

\subsection{Design and implementation}\label{subsec:qc-1}

\begin{figure}
    \begin{center}
    \includegraphics[width=\columnwidth, trim=2.5cm 8.2cm 11.15cm 9.3cm, clip]{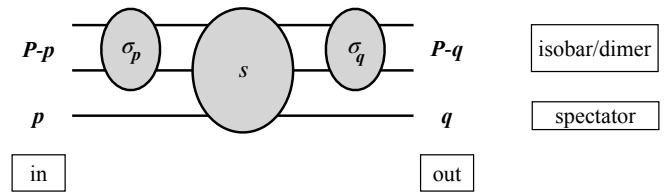}
    \end{center}
    \caption{\label{fig:kin}
    Schematic representation of a two-plus-one decomposition of the three-body system. $\bm{P}$, $\bm{p}/\bm{q}$ denote three-momenta of the three-body system and that of the in/outgoing spectators, respectively. Total three-body energy squared and the invariant mass of the two-body systems are denoted by $s$ and $\sigma_{\bm{p}/\bm{q}}$, respectively.
    }
\end{figure}

We avoid discussing the derivation of the FVU formalism here~\cite{Mai:2017bge,Mai:2018djl,Mai:2019fba,Culver:2019vvu}, but review the results and unify the notation. At its core, the condition derives from a relativistic unitary three-body scattering amplitude, which resolves the three-body dynamics as a cluster of a two-body (also related to `isobar' or `dimer' notation) state and a third particle -- a `spectator'~\cite{Mai:2017vot} (see also Refs.~\cite{Jackura:2018xnx,
Jackura:2020bsk, 
Mikhasenko:2019vhk
} for recent progress in this direction). The kinematic notation of such a configuration is depicted schematically in Fig.~\ref{fig:kin}. 
Unitarity constrains the correct interplay between the two- and three-body interactions accounting for on-shell configurations, which are the source of singularities in the finite-volume correlator. The net result is that in a finite volume  $\sqrt{s}$ is the energy corresponding to an interacting three-body state when
\begin{align}
\label{eq:qc}
 0 =
  \det\mathcal{Q}_{L\eta\tilde{\bm{P}}}(s):=&
  \det\Big[B_0(s) + C_0(s)-\\
  &E_{L\eta}\big( K^{-1}(s)/(32\pi)+\Sigma_{L\eta\tilde{\bm{P}}}(s)\big) \Big] 
  \,.\nonumber
\end{align}
The determinant is taken over the $\mathcal{S}_{L\eta}\times\mathcal{S}_{L\eta}$ space, referring to the momentum of the in- and outgoing spectator (third particle). Here, $\mathcal{S}_{L\eta}=\{2\pi/L(n_1,n_2,n_3/\eta)|(n_1,n_2,n_3)\in \mathds{Z}^3\}$ refers to the momentum configuration from elongated boxes used by GWUQCD and $\tilde{\bm{P}}$ is the total three-momentum of the system. The remaining building blocks of this condition ($\in \mathrm{Mat}^{\mathcal{S}_{L\eta}\times\mathcal{S}_{L\eta}}$) are
\vspace{-0.2cm}
\begin{itemize}

\item[\ding{100}] 
$[E_{L\eta}]_{\bm{p}\bm{q}}=\delta^{3}_{\bm{p}\bm{q}}2L^3\eta\sqrt{{\bm{p}}^2+m_\pi^2}\,.$

\item[\ding{100}]
$[B_0(s)]_{\bm{p}\bm{q}}^{-1}=
-2E_{{\* p}
+{\* q}}(\sqrt{s} -E_{{\* p}}
-E_{{\* q}}-E_{{\* p}
+{\* q}})\,.$\\[0.2cm]
This singular and non-diagonal matrix originates from the one-particle exchange contribution, and is a direct consequence of three-body unitarity in the infinite volume. Here and in the following $E_{\bm{x}}:=\sqrt{\bm{x}^2+m_\pi^2}$ denotes the on-shell energy of a pion with a momentum $\bm{x}$.

\item[\ding{100}]
$
[\Sigma_{L\eta\tilde{\bm{P}}}(s)]_{\bm{p}\bm{q}}
=\delta^{3}_{\bm{p}\bm{q}}
\frac{\sigma_{\bm{p}}}{L^3\eta} 
\sum_{\tilde{\bm{k}}\in \mathcal{S}_{L\eta}}
\frac{J\,\tilde J}{8E_{\bm{k}^*}^3(\sigma_{\bm{p}}-4E_{\bm{k}^*}^2)}\,.
$\\[0.2cm]
This singular, diagonal matrix accounts for the on-shell configurations in the two-body subsystem with total energy squared
$\sigma_{\bm{p}}(s)=(\sqrt{s}-E_{\bm{p}})^2-\bm{p}^2$. The equation above is derived in the Appendix with the final form given in Eq.~\eqref{eq:onesubtract}. Note that the summation is performed over $\tilde{\bm{k}}$ in the lattice frame, while the summands are expressed in terms of the $\bm{k}^*$ momenta in the two-body rest frame. The corresponding two-step boost reads\\[-0.5cm]
\begin{align*}
\nonumber
\colorbox{white}{
    \parbox{1.75cm}{
        \centering\parbox{1cm}{
            \centering lattice frame}}}
&\longrightarrow
\colorbox{white}{
    \parbox{1.75cm}{
        \centering \parbox{1.8cm}{
            \centering three-body rest frame}}}
&&\longrightarrow
\colorbox{white}{
    \parbox{1.75cm}{
        \centering \parbox{1.7cm}{
            \centering two-body rest frame}}}\\
\colorbox{white}{
    \parbox{1.75cm}{
        \centering $\tilde{\bm{k}}$}}
&\longmapsto
\colorbox{white}{
    \parbox{1.75cm}{
        \centering $\bm{k}(\tilde{\bm{k}},\tPvec,s)$}}
&&\longmapsto
\colorbox{white}{
    \parbox{1.75cm}{
        \centering $\bm{k}^*(\bm{k},\bm{p},s)$}.}
\end{align*}
The explicit formulas for the boosts and corresponding Jacobians $J$ and $\tilde J$ are collected in Appendix
\ref{sec:two-body-sums}.

\item[\ding{100}]
$[C_0(s)]_{\bm{p}\bm{q}}\,.$

This regular and in general non-diagonal matrix parametrizes the isobar-spectator interaction and, thus, also the three-body contact interaction via a non-trivial mapping discussed in the Sec.~\ref{subsec:qc-3}. In both cases, this contribution is not fixed and needs to be obtained from a fit to the lattice results as will be discussed in Sec.~\ref{subsec:qc-3} and \ref{sec:results}.
Note that the normalization used here differs slightly from previous work~\cite{Mai:2017bge, Mai:2017vot}.

\item[\ding{100}]
$[K^{-1}(s)]_{\bm{p}\bm{q}}=\delta^{3}_{\bm{p}\bm{q}}K^{-1}(\sigma_{\bm{p}})\,.$

This regular, diagonal matrix parametrizes the dynamics of the two-body sub-system. In that, it corresponds to the usual $K$-matrix as explained in detail in Appendix~\ref{app:mIAM}. Similarly to the three-body term $C_0$, this contribution needs to be determined from a fit to the lattice eigenvalues. Such a fit can take two-body, three-body, or both data types into account as will be discussed in Sec.~\ref{sec:results}.

\end{itemize}

Before proceeding with the discussion of the two- and three-body interaction terms $K^{-1}$ and $C_0$, we note a generic feature of the three-body quantization condition. In contrast to the well-established L\"uscher's method in the two-body case, the three-body quantization condition in Eq.~\eqref{eq:qc} emerges as a full-fledged determinant equation, even in the simplest case of identical particles in $S$-wave. Thus, it prevents any one-to-one mapping between infinite- and finite-volume quantities. Instead, even for the simplest three-body systems, one must fix the volume-independent quantities ($C_0$ and $K^{-1}$) from a fit to lattice data and then use those to evaluate the infinite-volume scattering amplitude. Note that this three-to-three amplitude provides only the three-body unitary final state interaction of production processes; to actually obtain mass-spectra or Dalitz plots as, e.g., in Ref.~\cite{Sadasivan:2020syi}, one needs additional information.

\subsection{Parametrization of the two-body input}
\label{subsec:qc-2}

Depending on the system at hand and the research objectives, various techniques in parametrizing the two-body dynamics ($K^{-1}$) may be more or less advantageous. For example, a model based on ChPT was used in Ref.~\cite{Mai:2018djl} to bridge the lattice results obtained at different pion mass.  Another approach, related to an effective range expansion was employed in Ref.~\cite{Blanton:2019vdk} allowing for a systematic extraction of threshold quantities.

With the lattice data spanning over large energy ranges and two different pion masses we proceed here with the path traced out in Ref.~\cite{Mai:2018djl}, and implement the modified Inverse Amplitude Method~\cite{Truong:1988zp, Dobado:1996ps, GomezNicola:2007qj, Hanhart:2008mx,GomezNicola:2007qj} (mIAM) into the three-body formalism. This is also motivated by our previous applications to the isoscalar channel~\cite{Doring:2016bdr,Guo:2018zss} and more recently cross-channel ($I=0,\,1,\,2$ $\pi\pi$ scattering) analysis~\cite{Mai:2019pqr} of GWUQCD lattice results~\cite{Guo:2018zss,Guo:2016zos,Culver:2019qtx} which is based on this approach.

\begin{figure}
    \includegraphics[width=\columnwidth, trim=0.9cm 9cm 9.3cm 9cm, clip]{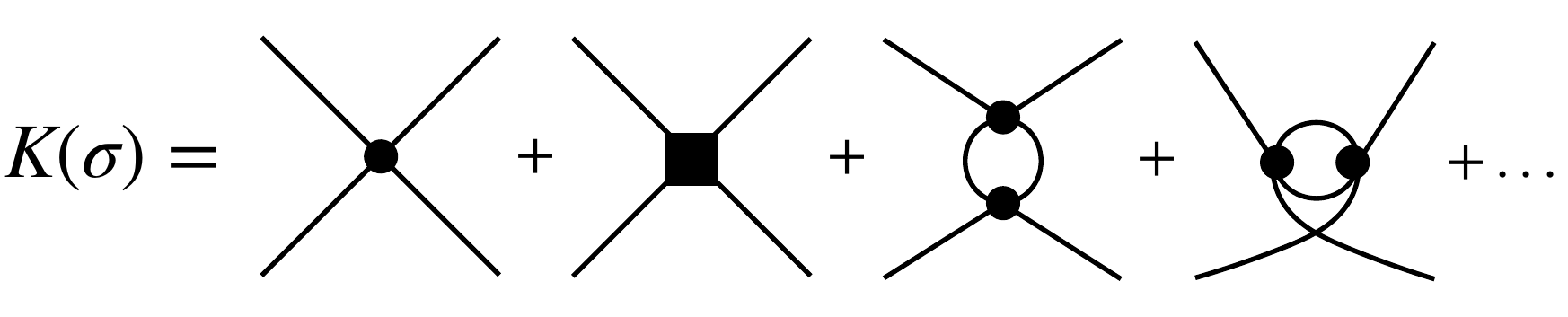}
    \caption{\label{fig:Kmat}
    Leading contributions to the $K$-matrix utilized for the two-body input of the quantization condition in Eq.~\eqref{eq:qc}. Diagrams represent scattering amplitudes obtained from ChPT with dots and squares denoting the leading and next-to-leading chiral order vertices, respectively (tadpole contributions not shown).}
\end{figure}

A practical implementation of this requires equating the two-body scattering amplitude used in the infinite volume analog of Eq.~\eqref{eq:qc} to the mIAM amplitude~\cite{Hanhart:2008mx}. This procedure has been developed in Ref.~\cite{Mai:2018djl}, but is improved in the current study. In particular, the matching procedure is now exact and without the need of a redundant regulator. Details can be found in Appendices~\ref{app:mIAM} and \ref{appsec:newtwo} leading to
\begin{align}
\label{eq:K-mat}
    K^{-1}(\sigma)=
    \frac{T_{2}^{0}(\sigma)-\bar T_{4}^{0}(\sigma)+A(\sigma)}{\left(T_{2}^{0}\left(\sigma\right)\right)^2}
\end{align}
 (see Eq.~\eqref{app:KMatrix} there).
Here, $T^\ell_n$ denotes the chiral $\pi\pi$ amplitude of order $n$ projected to the isospin $I=2$, angular momentum $\ell=0$ partial wave. The barred symbol denotes a reduced amplitude, where the $s$-channel one-loop diagram is subtracted. This algebraic re-shuffling ensures that the above expression indeed contains only quantities which do not go on-shell in the physical region. Similarly, $A(\sigma)$ is a function of the chiral amplitudes, introduced in Ref.~\cite{Hanhart:2008mx} to improve the analytic behavior of the (infinite volume) scattering amplitude below threshold $0<\sqrt{\sigma}<2m_\pi$.
In particular, adding $A(\sigma)$ prevents the two-body scattering amplitude ($\sim 1/(T_2-T_4)$) from diverging around the Adler zero ($\sigma \Leftrightarrow T_{22,\infty}(\sigma)=0$, where $T_{22,\infty}$ is the $S$-wave scattering amplitude). Recall that the presence of the Adler zero itself is demanded by chiral symmetry.
The diagrammatic representation of the leading contributions is given in Fig.~\ref{fig:Kmat}, while formulas are provided for convenience in Appendix~\ref{app:mIAM}.

Note that the two-body subthreshold amplitude enters the three-body quantization condition Eq.~\eqref{eq:qc} since for some $\tilde{\bm{p}}$,   $\sigma_{\bm{p}(\tilde{\bm{p}},\tilde{\bm{P}},s)}<4m_\pi^2$
for $\tilde{\bm{p}}\in\mathcal{S}_{L\eta}$. The mIAM approximates the left-hand cut by the next-to leading chiral order. Thus, it is expected that at too low invariant masses the latter model is a poor approximation of reality. To account for this, we have explored various possibilities, such as employing smooth form-factors, etc.. We found that fixing $K^{-1}(\sigma)\mapsto K^{-1}(\sigma_0)$ for $\sigma<\sigma_0=3m_\pi^2$ is sufficient. Variating the matching point in a reasonable range leads to uncertainties orders of magnitude smaller than other effects, see Ref.~\cite{Alexandru:2020xqf}. Also, changes in sub-threshold behavior are supposed to be absorbed in three-body contact terms.

Overall, the expression in Eq.~\eqref{eq:K-mat} is a regular,  volume-independent function of two-body energy $\sigma$, ensuring in infinite volume an exact matching of the two-body dynamics of the mIAM approach. In that, Eq.~\eqref{eq:K-mat} depends on four renormalized low-energy constants $\{l^r_1, l^r_2, l^r_3, l^r_4\}$, see Ref.~\cite{Gasser:1983yg} for explicit formulas relating those to Lagrangian constants. As it has been shown in previous studies of two-body lattice results~\cite{Acharya:2015pya,Doring:2016bdr}, $l_3^r$ and $l_4^r$ contribute weakly to the two-body $\pi\pi$ dynamics. This is simply due to the fact that their appearance is solely rooted in the procedure of replacing the Lagrangian quantities -- pion decay constant in the chiral limit and leading order pion mass -- by their physical/lattice values. Thus, we simply fix the symmetry breaking LECs $\{l_3^r=8.94\cdot 10^{-6},\, l_4^r=9.05\cdot10^{-3}\}$ to their typical values~\cite{Mai:2019pqr,Aoki:2019cca} at the regularization scale of $\mu=770~\rm MeV$. Another possibility is to directly use  both quantities as fit parameters as introduced in Ref.~\cite{Fischer:2020fvl}. This has an advantage of avoiding the scale setting discussion, which is beyond the current stage of three-hadron spectroscopy from QCD.

\subsection{Parametrization of the three-body input}
\label{subsec:qc-3}

\begin{figure*}[t]
    \includegraphics[width=0.9\linewidth,trim=0 9cm 2cm 9cm, clip]{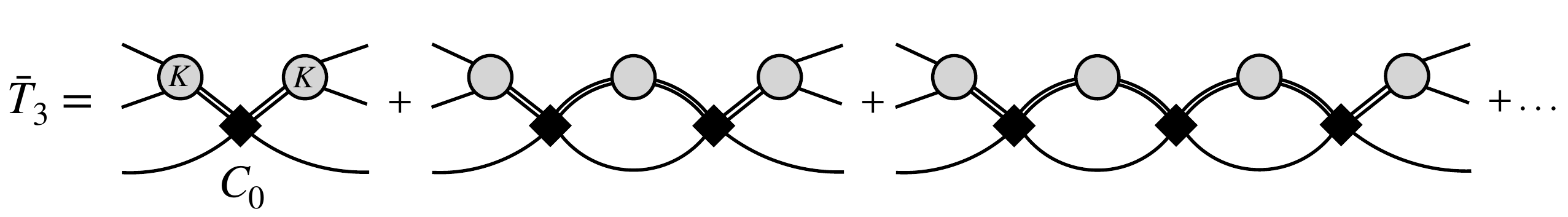}
    
    \noindent\rule{\linewidth}{0.4pt}\\[0.5cm]
    
    \includegraphics[width=0.31\linewidth, trim=0 0 3cm 0, clip]{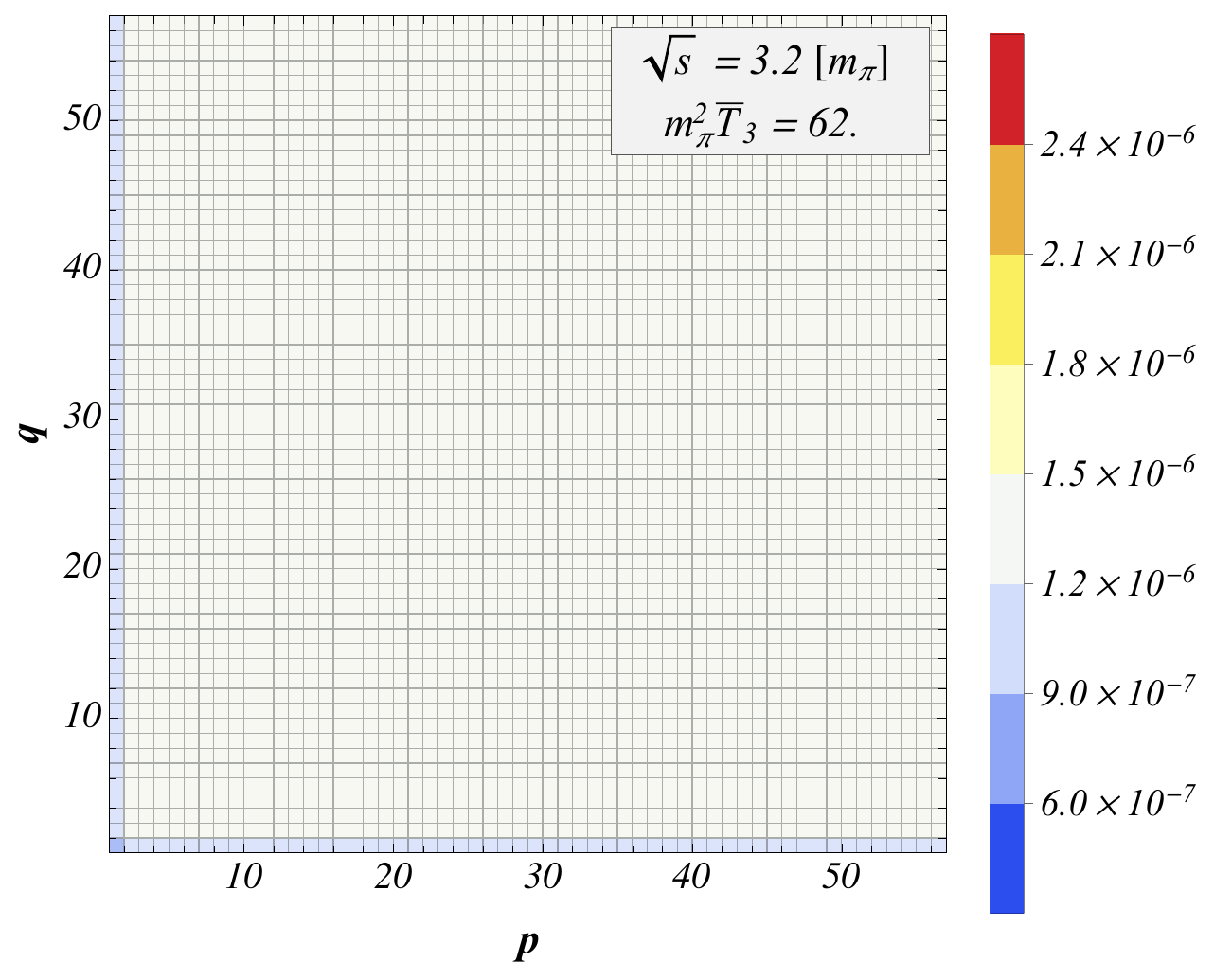}~~~~~
    \includegraphics[width=0.29\linewidth, trim=0.5cm 0 3cm 0, clip]{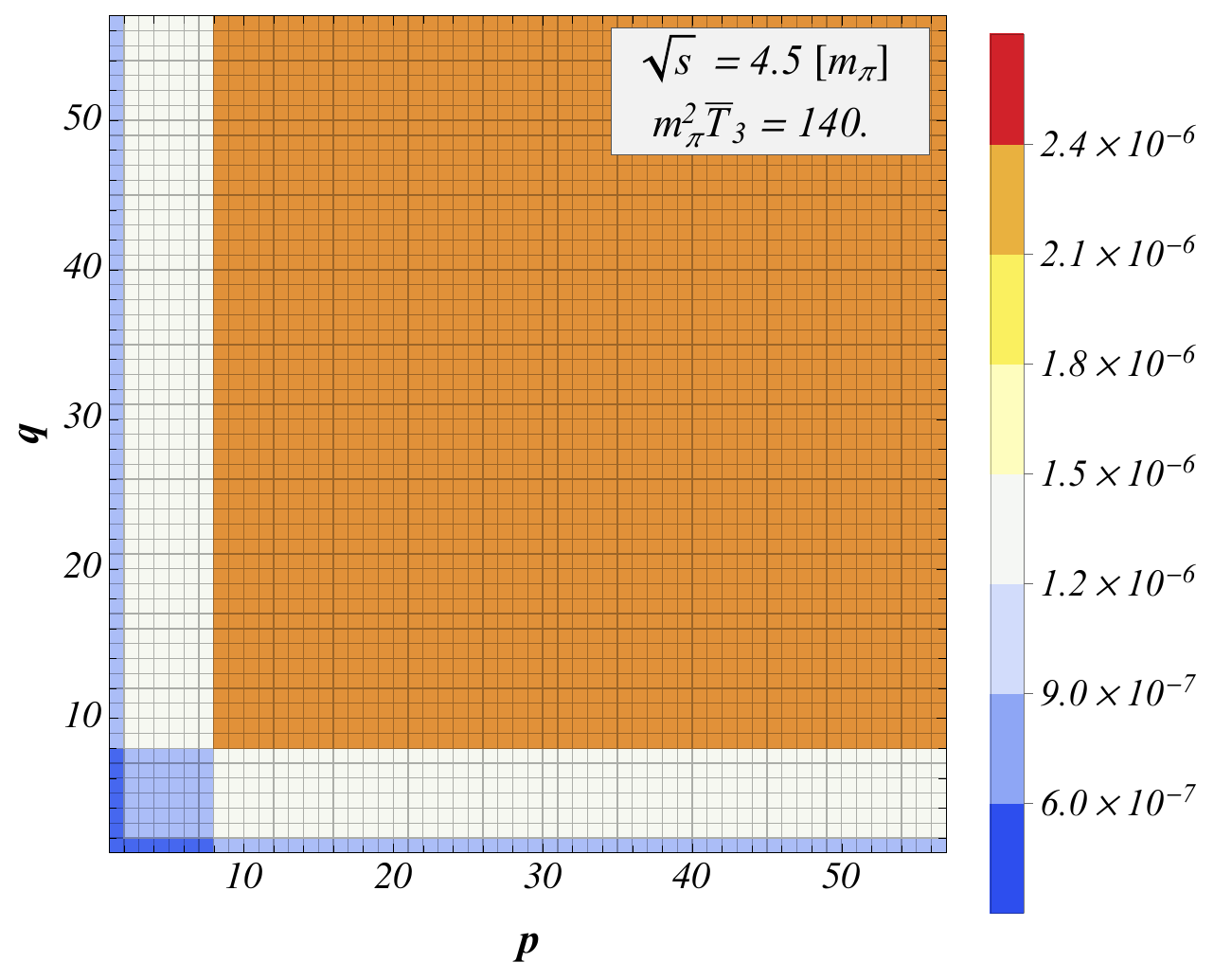}~~~~~
    \includegraphics[width=0.38\linewidth, trim=0.5cm 0 0 0, clip]{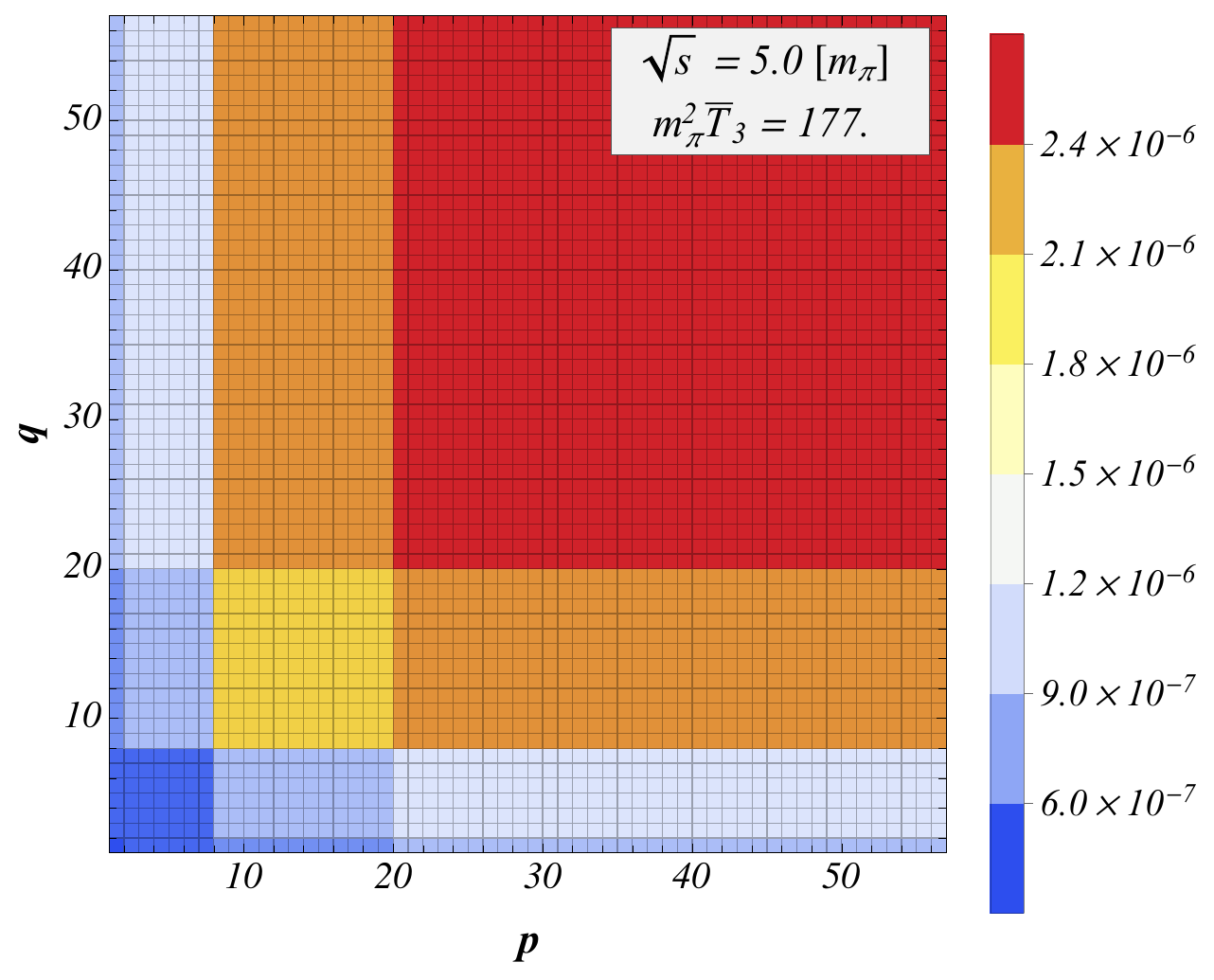}
    \caption{
    \label{fig:T3bar}
    {\bf Top:} Connection between the contact three-to-three ($\bar T_3$) and isobar-spectator interaction ($C_0$). Two-body dynamics is encoded in the $K$-matrix, which does not contribute to  divergences.
    {\bf Bottom:} Visualisation of $C_0$ (in ${\rm MeV}^{-2}$) from Eq.~\eqref{eq:C0-to-T3bar} for $\bar T_3$ from leading order ChPT at relevant values of total three-body energy $\sqrt{s}$. In- and  outgoing spectator momenta $\bm{p}$ and $\bm{q}$ are given by their index in the set $(2\pi)/(L)\{(0,0,0),...,(\pm2,\pm1,0)\}$, ordered by magnitude.
    }
\end{figure*}

The implementation of the three-body input into the quantization condition Eq.~\eqref{eq:qc} is much less explored than that of the two-body. One of the goals of this paper is to improve on this, in exploring various ways of parametrizing three-body dynamics theoretically and in a practical application to the lattice data.

In the three-body finite-volume formalism (FVU)~\cite{Mai:2017bge}, 
the three-body dynamics is included via a regular (real-valued, infinite volume) function --  the isobar-spectator contact term $C \sim C_0$. This term is not fixed by unitarity and, other than demanded by data, there are no constrains on its functional dependence with respect to energies and masses. The RFT formalism~\cite{Hansen:2014eka} on the other hand, relies on the diagrammatic counting of the on-shell configurations in derivation of the three-body quantization condition. Thus, it includes by construction a regular three-body contact term ($K_{3,\rm df}$), which again needs to be determined from a fit to the data.

Interestingly, in the pioneering application to the lattice $3\pi^+$ finite-volume spectrum~\cite{Mai:2018djl}, a simplistic fit\footnote{Note that projection to shells (sets of momenta related by octahedral symmetry transformation), was employed in this work, which is not part of the more general form of Eq.~\eqref{eq:qc}. There the employed fit form was chosen as $C_0=\mathds{1}\cdot c_0 (m_\pi/m_\pi^{\rm phys})^2$.} to the NPLQCD data~\cite{Detmold:2008fn} led to $c_0=(0.3\pm2.3)\times 10^{-6}~[{\rm MeV}^{-2}]$. Later, an analysis of more recent data at one pion mass~\cite{Horz:2019rrn} was performed within both RFT~\cite{Blanton:2019vdk} and FVU~\cite{Mai:2019fba}. In that, the FVU formalism with vanishing three-body term $C_0$ led to a $\chi^2_\text{dof}\approx 0.9$ prediction of three-body energies, while a non-zero $m_\pi^2K_{3,\rm df}=270(160)$ was obtained in the RFT approach by fitting two- and three-body lattice energies. The apparent discrepancy between these results was raised in the community, but, so far, not resolved quantitatively.

To find an algebraic connection between the three-to-three contact term ($\bar T_3$) and the isobar-spectator coupling $C_0$ we utilize the language of Ref.~\cite{Mai:2017bge,Mai:2017vot} in matrix notation for convenience. There, a fully connected three-body amplitude takes the form
\begin{align}
    T_3=\frac{1}{3!}\langle v \tau T_{\rm IS}\tau v\rangle\,
    \label{eq:teone}
\end{align}
for $T_{\rm IS}$, $\tau$ and $v$ denoting isobar-spectator amplitude, isobar propagator and its coupling to asymptotically stable states, respectively. Explicit definitions of the latter do not matter for the derivation but can be found in Refs.~\cite{Mai:2017bge,Mai:2017vot} and Appendix~\ref{appsec:newtwo}.
For simplicity, momentum and energy labels are omitted. Symmetrization over external momenta is taken into account by $\langle...\rangle$. For three degenerate scalars in (relative) $S$-wave this leads to
\begin{align}
    T_3=\frac{3}{2} v\tau \frac{\mathds{1}}{
    \mathds{1}+v(B_0+C_0)vE_{L\eta}^{-1}\tau}
    v(B_0+C_0)v\tau v\,.
\end{align}
Taking now the limit $m_\pi\to\infty$ for all intermediate pions provides an expression equivalent to a contact term ($\bar T_3$)
\begin{align}
\label{eq:T3bar-to-C0}
  \bar T_3=
  \frac{3}{2}
  \left(\frac{K^{-1}}{32\pi}\right)^{-1}
  \frac{C_0}{
  \mathds{1}-C_0E_{L\eta}^{-1}
  \left(\frac{K^{-1}}{32\pi}\right)^{-1}
  }
  \left(\frac{K^{-1}}{32\pi}\right)^{-1}
  \,,
\end{align}
or equivalently
\begin{align}
\label{eq:C0-to-T3bar}
  C_0=
  \left(\frac{K^{-1}}{32\pi}\right)
  \Big(\frac{2}{3}\bar T_3\Big)
  \left(\frac{K^{-1}}{32\pi}\right)
  \frac{\mathds{1}}{
  \mathds{1}+E_{L\eta}^{-1}\big(\frac{2}{3}\bar T_3\big)\left(\frac{K^{-1}}{32\pi}\right)
  }\,,
\end{align}
where $\bar T_3$ denotes a real three-body contact term. This equation is schematically illustrated in Fig.~\ref{fig:T3bar} to the top. Note again that appearance of factors $(32\pi)$ is caused by the fact that we are working in the plane wave basis, see Appendix~\ref{app:mIAM} for more details. This relation is of the same form as the relation between the $K_{\rm 3,df}$ and $C_0$ term, which can be obtained from a matching of FVU and RFT formalisms, see for example Refs.~\cite{Hansen:2019nir,Rusetsky:2019gyk}. Also, as noted there, it implies that in general an isotropic $\bar T_3$ leads to anisotropic $C_0$ and vice versa.

To expand on this further, we consider the following example. Chiral perturbation theory at leading chiral order (this was used for the RFT formalism in Ref.~\cite{Blanton:2019vdk}) yields for our formalism a three-to-three contact term of the form
\begin{align} \label{eq:ChPTpred}
  \bar T_3=\frac{1}{27f_\pi^4}\left(4s-9m_\pi^2\right)\,.
\end{align}
Using now Eq.~\eqref{eq:C0-to-T3bar} with the $K$-matrix from Eq.~\eqref{eq:K-mat} we obtain a prediction for the isobar-spectator interaction $C_0(\sqrt{s},\bm{p},\bm{q})$, for the momenta belonging to the in/outgoing spectators. The resulting symmetric (in $\bm{p},\bm{q}$) isobar-spectator contact interaction is depicted for several values of total three-body energy $3m_\pi<\sqrt{s}<5 m_\pi$ in Fig.~\ref{fig:T3bar}. We observe that several orders of magnitude difference between the overall scales of the $\bar T_3$ and $C_0$ (e.g. $C_0\approx 10^{-6}~{\rm MeV}^{-2}\leftrightarrow m_\pi \bar T_3\approx 10^2$) occurs naturally, connecting the results of Refs.~\cite{Blanton:2019vdk,Mai:2019fba}.

\subsection{Three-body state energies}
\label{sec:fits}

Given parameterizations of the two- and three-body interactions, the finite-volume spectrum can be determined by searching for energies at which Eq.~\eqref{eq:qc} is satisfied. 
To find the energies associated with a particular irrep $\Lambda$ of the symmetry group $G$, we first block-diagonalize the matrix~${\cal Q}$
\begin{equation}
    \begin{aligned}
    \mathcal{Q} &= \diag({\cal Q}_{\Lambda_1},{\cal Q}_{\Lambda_2},\dots),\\
    \Rightarrow \det{\cal Q} &= \prod_i \det{\cal Q}_{\Lambda_i}.
    \end{aligned}
\end{equation}
The determinant of ${\cal Q}_\Lambda$ block can then be scanned separately in $s$ for the zeros which determine the energies finite-volume states for that irrep.

To block-diaganolize the matrix ${\cal Q}$ we need to change to an appropriate basis. The change of basis from the plane-wave basis in Eq.~\eqref{eq:qc} can be done by constructing projectors for the row $\lambda$ of the irrep $\Lambda$ of the group $G$:
\begin{equation}\label{eq:projector}
    P_{\Lambda\lambda} = \frac{d_\Lambda}{|G|}\sum_{g\in G} \Gamma_{\lambda\lambda}^{\Lambda}(R(g)) U(R(g))^\intercal,
\end{equation}
where $d_\Lambda$ is the dimension of the irrep $\Lambda$, $\Gamma_{\lambda\lambda}^\Lambda(R)$ is the representation matrix of the group element $g$ in row $\lambda$ of the irrep $\Lambda$, $R(g)$ is the rotation corresponding to $g$, and $U(R(g))$ is the unitary transformation of $R$ on the plane-wave space. 

We must truncate the plane-wave
space ${\cal S}_{L\eta}$ to include only momenta below a
sufficiently large threshold, $p_\text{max}\approx 1\GeV$. 
In general the values of the parameters would need to be adjusted based on the choice of the cutoff $p_\text{max}$. In our case we found that
there is a very mild dependence on the choice of $p_\text{max}$, for the energy levels predicted from the quantization condition, well below the stochastic errors in the lattice data.
By construction
this truncated space is invariant under the symmetry transformations, since the momenta magnitudes are invariant
under rotations. For each
irrep~$\Lambda$ we project the entire plane-wave basis using
the projector above and the dimensionality of this space
represents the number of independent multiplets associated
with $\Lambda$ that appear in the selected plane-wave basis.
The restriction of ${\cal Q}$ to this subspace is ${\cal Q}_\Lambda$ which has zeros corresponding to energies in the irrep~$\Lambda$.

\begin{figure}
    \centering
     \includegraphics[width=\columnwidth, trim = 0.2cm 1.2cm 1.5cm 0.8cm]{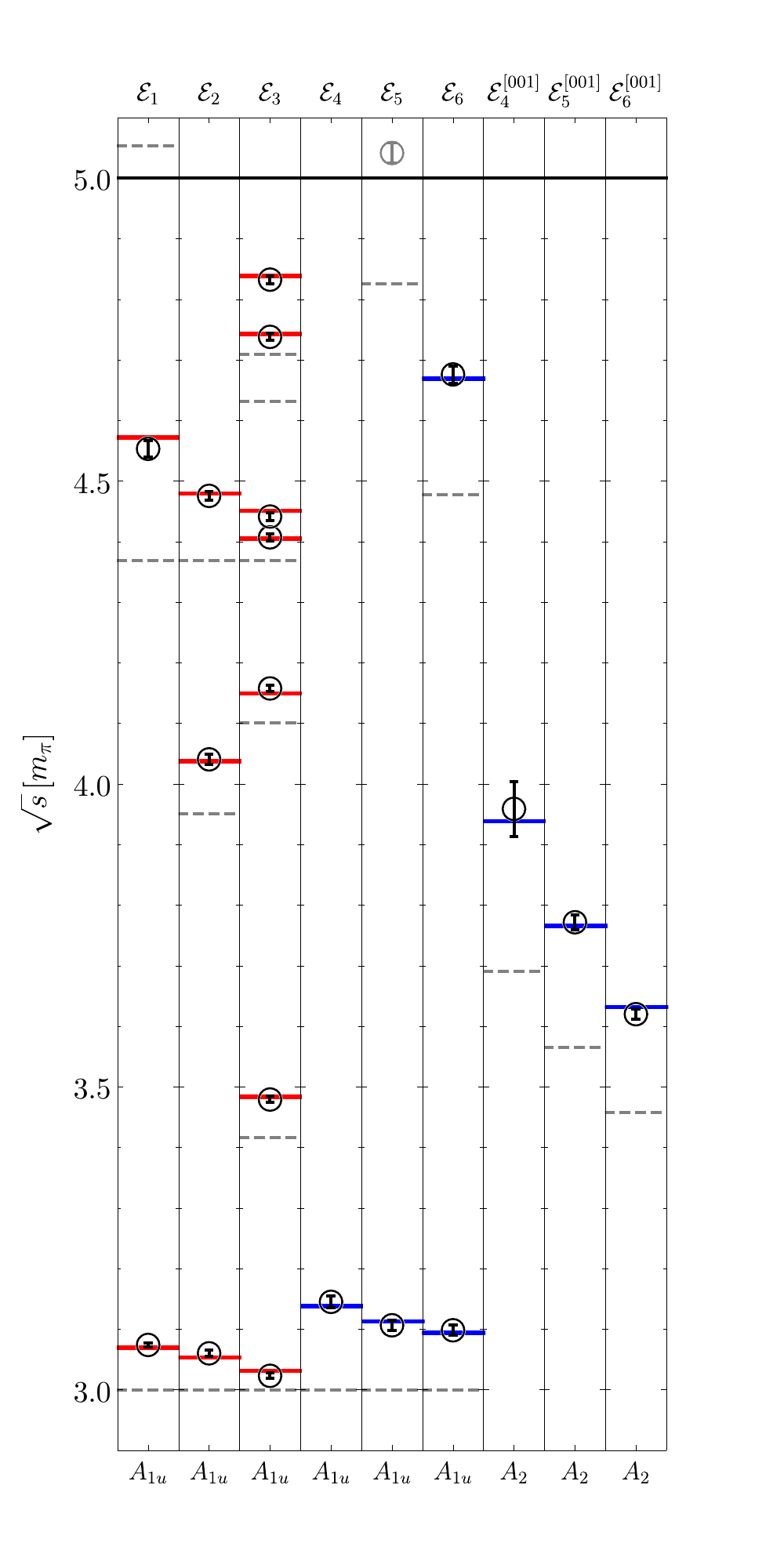}
     \caption{Finite-volume center-of-mass $\pi^+\pi^+\pi^+$ energies for $m_{\pi}=315\MeV$ ($\mathcal{E}_{1,2,3}$)   and $m_{\pi}=220\MeV$ ($\mathcal{E}_{4,5,6}$). For each pion mass there is one cubic box (${\cal E}_{1,4}$) and two elongated boxes (${\cal E}_{2,3,5,6}$). Columns distinguish different irreps of the rotational symmetry group containing energies below the inelastic threshold, $5m_\pi$ (solid black line). The data points are the LQCD energy levels with error bars inside of the circles. The dashed lines are the non-interacting energy levels. Boosted frames with non-zero total momentum are denoted by the superscript $[001]$ indicating a single unit of momentum in the elongation ($z$) direction. The solid lines represent the predicted central values of the spectrum from FVU, after fitting $\overline{t}_0$, $\overline{t}_1$, $l^r_1$, and $l^r_2$, for $m_{\pi}=315~\MeV$ (red) and  $m_{\pi}=220~\MeV$ (blue) separately.
     }
\label{fig:spectrum}
\end{figure}

\begin{figure*}[t]
    \centering
    \includegraphics[height=5.3cm, trim=0 0 0.3cm 0, clip]{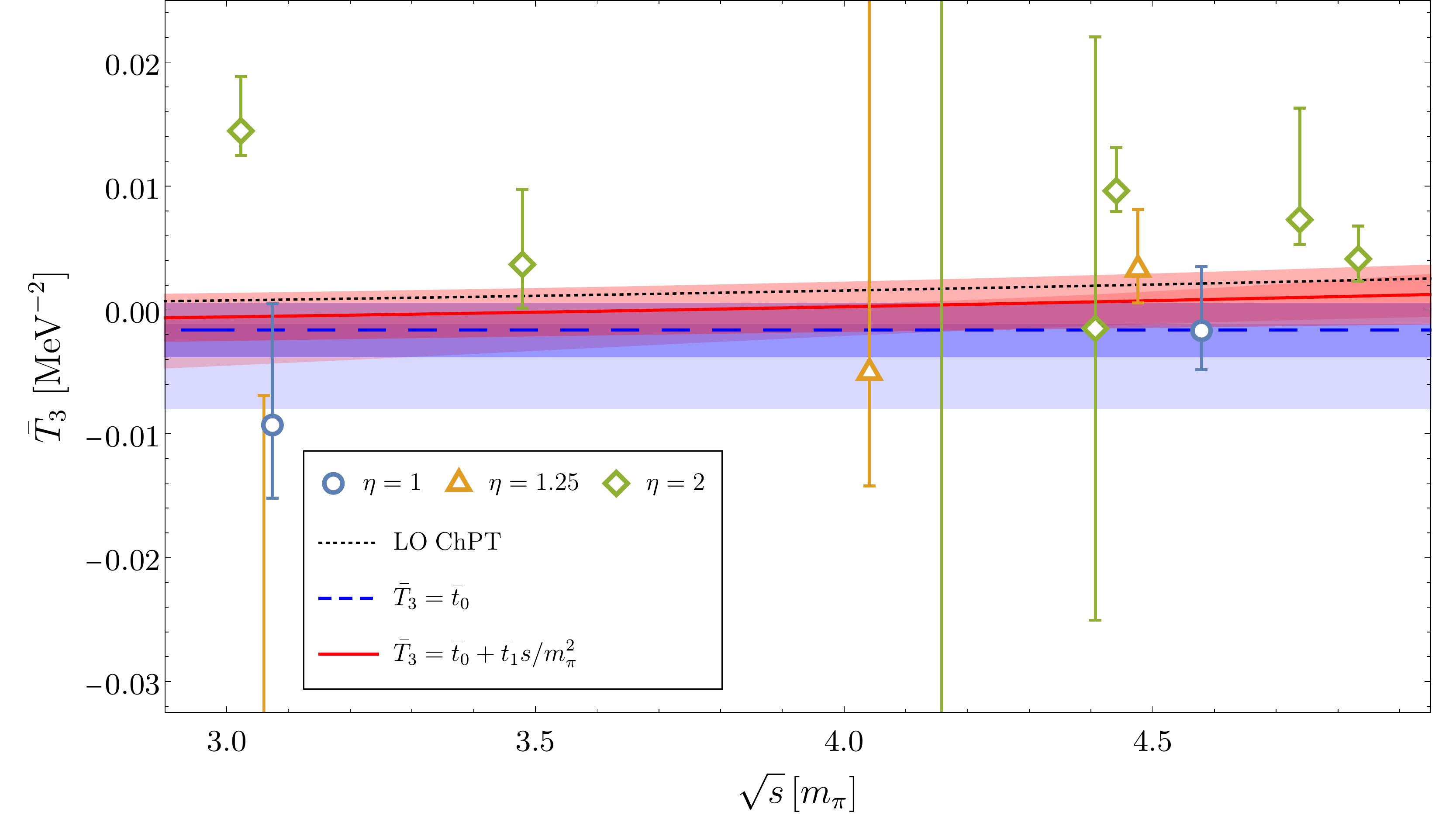}
    \includegraphics[height=5.3cm, trim=3.7cm 0 0.25cm 0, clip]{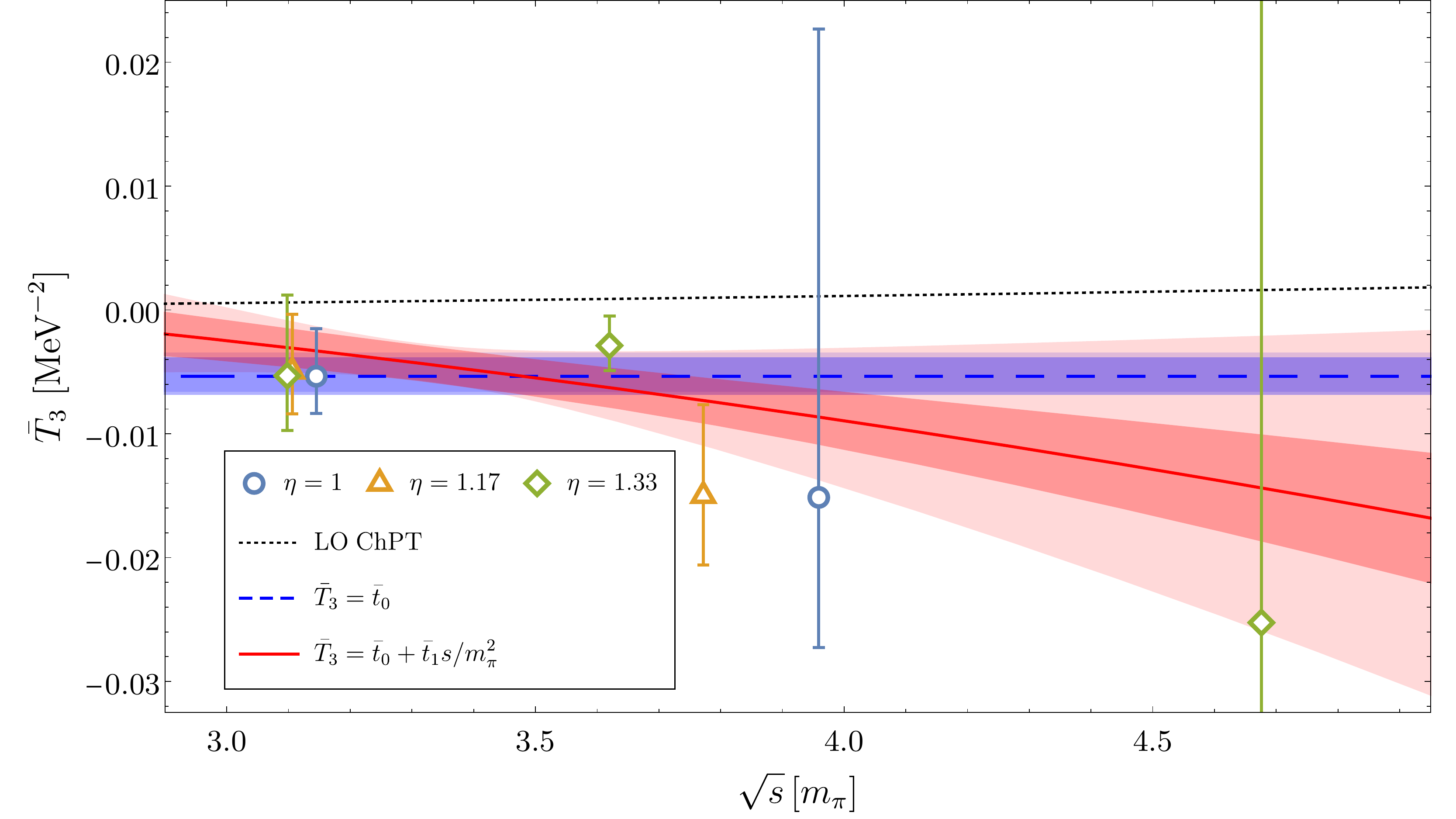}
    \caption{
   The three-body contact term $\bar{T}_3$ defined via Eq.~\protect\eqref{eq:C0-to-T3bar} as a function of three-body energy $\sqrt{s}$ for $m_\pi=315$~MeV (left) and $m_\pi=220$~MeV (right). The data with error bars show the pointwise determination of $\bar T_3$. The three lines correspond to constant, linear and ChPT energy dependence. Dark bands indicate $1\sigma$ uncertainties for fits of the three-body data using fixed mIAM LECs from Ref.~\cite{Mai:2019pqr}. Light bands indicate the $1\sigma$ uncertainties from fits to the three-body data including $l_1^r$ and $l_2^r$ as fit parameters and priors from the cross-channel two-pion fits as described in Sec.~\ref{sec:results}.
   }
    \label{fig:lightTomo}
\end{figure*}

\section{Results}\label{sec:results}

In this section we present the results for the three-body terms as extracted from fitting the finite-volume spectrum. For sake of clarity we will first discuss the extraction of the three-body terms when the two-body dynamics is fixed by the mIAM parameterization with LECs determined from fitting our lattice two-pion spectrum in all isospin channels~\cite{Mai:2019fba}. This model provides a good description of our two-pion spectrum across all channels for the two quark masses we studied. For the three-body terms we parameterize $\bar{T}_3$, using no energy dependence: $\bar{T}_3(s) = \overline{t}_0$, or $\bar{T}_3(s) = \overline{t}_0 +  \overline{t}_1 s/m_\pi^2$, a linear function in $s$.
The second parametrization is consistent with the leading order prediction from ChPT in Eq.~\eqref{eq:ChPTpred}.

The data points included in these fits are all three-body energies below the inelastic threshold in
the irreducible representations sensitive to the s-wave three-body terms. The relevant representations are $A_{1u}$ for the states at rest and $A_2$ for the moving states. The total number of points is 12 (2, 3, and 7 for ensembles ${\cal E}_1$, ${\cal E}_2$, and ${\cal E}_3$ respectively) for the heavy quark mass and 7 (2, 2, and 3 for ensembles ${\cal E}_4$, ${\cal E}_5$, and ${\cal E}_6$ respectively) for the light quark mass. These energy levels are plotted with errors in Fig.~\ref{fig:spectrum}.

All fits to the three-body energy levels are performed separately for the two quark masses since our parametrization for the three-body term does not constrain the quark mass dependence. We perform various fits, keeping some of the fit parameters fixed and varying others as shown in the Table~\ref{tab:fitsT3lecs}. Below we discuss these results.

To get a sense whether these parametrizations are reasonable, we first extracted $\bar{T}_3(s)$ by fitting each individual lattice energy level alone.  This offers a rough profile of $\bar{T}_3$ as a function of energy. To stabilize these fits we fix the LECs to the values extracted in our cross-channel two-pion study~\cite{Mai:2019fba}. In Fig.~\ref{fig:lightTomo} these results are represented by the data points, for the heavy and light quark masses. We note that while the errors are large the results support $\bar{T}_3(s)$ being (weakly) dependent on energy. For the heavier quark mass the data is consistent with a constant parameterization for $\bar{T}_3$, close to zero, whereas for the lighter mass we see a statistically significant fall in the data with energy. Note that the magnitude of $\bar{T}_3$ in absolute terms is not that different, but the energy dependence is more pronounced for the lighter mass. Additionally, for the light mass $\bar{T}_3$ is well constrained to be non-zero away from threshold. We also present in these plots the expectation from the leading order ChPT, plotted with a dashed line. At the light mass there is some tension with the leading order ChPT prediction, in particular at higher energies. Also, the energy dependent term has opposite sign to the ChPT result. This is similar to the tension in the energy dependence of the RFT three-body term in Refs.~\cite{Blanton:2019vdk,Fischer:2020jzp}.

\begin{table}[b]  
\caption{Covariance matrix $\Sigma_{ij}$ for the LECs $l_1^r$, $l_2^r$ as determined in Ref.~\cite{Mai:2019fba}.}
\label{tab:crossLEC}  
\begin{tabular*}{\columnwidth}{@{\extracolsep{\fill}}*{5}{>{$}c<{$}}@{}}
\toprule
 & \hat{l}_i & \phantom{andrei} & \multicolumn{2}{c}{$\Sigma_{ij}/(\sigma_i\sigma_j)$} \\
\midrule
l^r_1 & -0.00407(12) && 1 & 0.744 \\
l^r_2 & \,\,\,\,\,0.00514(20) &&  & 1 \\
\bottomrule
\end{tabular*}
\end{table}

\begin{table*}[t]
    \resizebox{.5\textwidth}{!}{
    \begin{tabular}{@{}*{14}{>{$}l<{$}}@{}}  
    \toprule
    \overline{t}_0\cdot10^3{\rm MeV}^2 & \overline{t}_1\cdot10^3{\rm MeV}^2 & l_1^r\cdot10^3 & l_2^r\cdot10^3 & \chi^2_{I=3} &+&\chi^2_{\rm priors} & \chi^2_{\rm dof} \\
    \midrule
    \bf\hphantom+0.0 & \bf \hphantom+0.0 & \bf-4.07 & \bf+5.14 & 27.39&&0 &  2.28  \\
    -1.7(2.3) & \bf \hphantom+0.0 & \bf -4.07 & \bf+5.14 &  25.95&&0 & 2.36 \\
    -1.7(2.1) & +0.12(10) & \bf -4.07 & \bf+5.14 & 25.76&&0 & 2.58 \\
    \bf\hphantom+0.0 & \bf\hphantom+0.0 & -4.034(85) & +5.21(12) & 27.08 &&0.14  & 2.27 \\
    -4.5(3.5) & \bf\hphantom+0.0 & -3.947(98) & +5.39(15) & 21.41 & &1.52 & 2.08 \\
    -4.5(3.5) & +0.24(18) & -4.038(93) & +5.20(14)~~~~ & 22.33 & &0.10  & 2.24 \\
    \midrule
    \bf \hphantom+0.0 & \bf \hphantom+0.0 & -4.3(1.5) & +5.42(83) & 27.11 && 0.03^* &
    
    \\
    \bottomrule
    \end{tabular}
    }
    \resizebox{.49\textwidth}{!}{
    \begin{tabular}{@{}*{14}{>{$}l<{$}}@{}}  
    \toprule
    \overline{t}_0\cdot10^3{\rm MeV}^2 & \overline{t}_1\cdot10^3{\rm MeV}^2 & l_1\cdot10^3 & l_2^r\cdot10^3 & \chi^2_{I=3} & +&\chi^2_{\rm priors} & \chi^2_{\rm dof} \\
    \midrule
    \bf\hphantom+0.0 & \bf\hphantom+0.0 & \bf -4.07 & \bf+5.14 &  33.78 &&0& 4.83 \\
    -5.3(1.5) & \bf\hphantom+0.0 &\bf-4.07 & \bf+5.14 &  15.10 &&0& 2.52 \\
    +5.8(4.5) & -0.92(38) & \bf-4.07 & \bf+5.14 &  10.41 &&0& 2.08 \\ 
    \bf \hphantom+0.0 & \bf\hphantom+0.0 & -4.27(12) & +4.75(20) & 25.80     & &3.79 & 4.23 \\
    -5.0(1.6) & \bf\hphantom+0.0 & -4.12(12) & +5.03(20) & 14.50 && 0.30 & 2.47 \\
    +6(12) & -0.9(1.1) & -4.10(13) & +5.09(22)~~~~ & 10.27 & &0.07 & 2.07 \\
    \midrule
    \bf\hphantom+0.0 & \bf\hphantom+0.0 & -5.1(1.8) & +3.2(1.1)  & 7.11 && 0.11^* & 
     
    \\
    \bottomrule
    \end{tabular}
    }
    \caption{
Fit results for $\bar{T}_3$ and LECs including $I=3$ $\pi\pi\pi$ energies only for $m_\pi=315$~MeV (left) and $m_\pi=220$~MeV (right). Bold font indicates parameters fixed to values from Ref.~\cite{Mai:2019pqr}, others are left as free parameters of the fit. The final row in each table is for a fit using {\it relaxed} priors, as indicated by an asterisk and described in the text.}
\label{tab:fitsT3lecs}
\end{table*}

The simultaneous fit to all the energy levels are
indicated with color bands in  Fig.~\ref{fig:lightTomo}. For these fits we allowed the LECs to vary but constrained their variation using priors based on the probability distribution that was determined in our cross-channel two-pion data fits.
We prefer this strategy over the simultaneous fit of the two- and three- body energies, since it makes the results for the three-body fit easier to analyze by isolating the two-body contribution.
These priors are included by augmenting the $\chi^2$-function that
is minimized by the fitter:
\beq
\chi^2_\text{aug} = 
\chi^2_0 + \underbrace{\delta l_i (\Sigma^{-1})_{ij} \delta l_j}_{\chi^2_\text{priors}}\,,
\eeq
where $\delta l_i:= l_i^r - \hat{l}_i$, with $\hat{l}$ being the
values of the LECs as determined in Ref.~\cite{Mai:2019fba} and $\Sigma_{ij}$ their covariance matrix. Note that since the LECs variations tend to be strongly correlated, in order to include these priors properly we need to consider the full correlation matrix $\Sigma$, not just the diagonal terms associated with individual LECs error estimates. In Ref.~\cite{Mai:2019fba} we only reported the errors on the LECs. We include the estimate for the covariance matrix in Table~\ref{tab:crossLEC}.

The constant fit works reasonably well for the heavy quark mass ensembles and is consistent with the linear fit in $s$. For the lighter quark mass the linear and constant fit are quite different and the linear term is needed to describe the data well. The values extracted for the fit constants and the $\chi^2$ value for the fit are listed in Table~\ref{tab:fitsT3lecs}.

We also indicate in the figures with narrower bands the fit results for $\{\overline{t}_0,\overline{t}_1\}$ when the LECs are not allowed to vary, corresponding to the second and third rows in Table~\ref{tab:fitsT3lecs}. We see that while the central values are almost the same, the error bands are almost doubled when we allow the LECs to vary. This shows that even with small error bars~(few percent level) in the two-body parametrization, there is a large impact on the error of the three-body terms in this channel. This is partly due to the smallness of three-body terms. The smallness is unsurprising since the three-body effects are suppressed by an additional volume factor relative to two-body interactions.

In terms of fit quality, we note that for both masses, the fits for the linear form with varying LECs produce a $\chi^2$ per degree of freedom around~$2$. This indicates that there is a slight tension between the data and the parametrization used here. At this point it is not easy to determine whether this is the result of the quality of the lattice data or due to the lack of flexibility in the fit form used for the three-body terms. We note that since the three-body predictions are sensitive to the two-body inputs, some of this tension might have as a source small discrepancies in the two-body amplitudes used in the quantization conditions. We note that when analyzing two-meson energy levels using mIAM framework, a similar level of agreement between data and predictions was found~\cite{Mai:2019pqr}.

\begin{figure*}[t]
    \centering
    \includegraphics[width=0.495\textwidth,trim=0 0 0cm 0,clip]{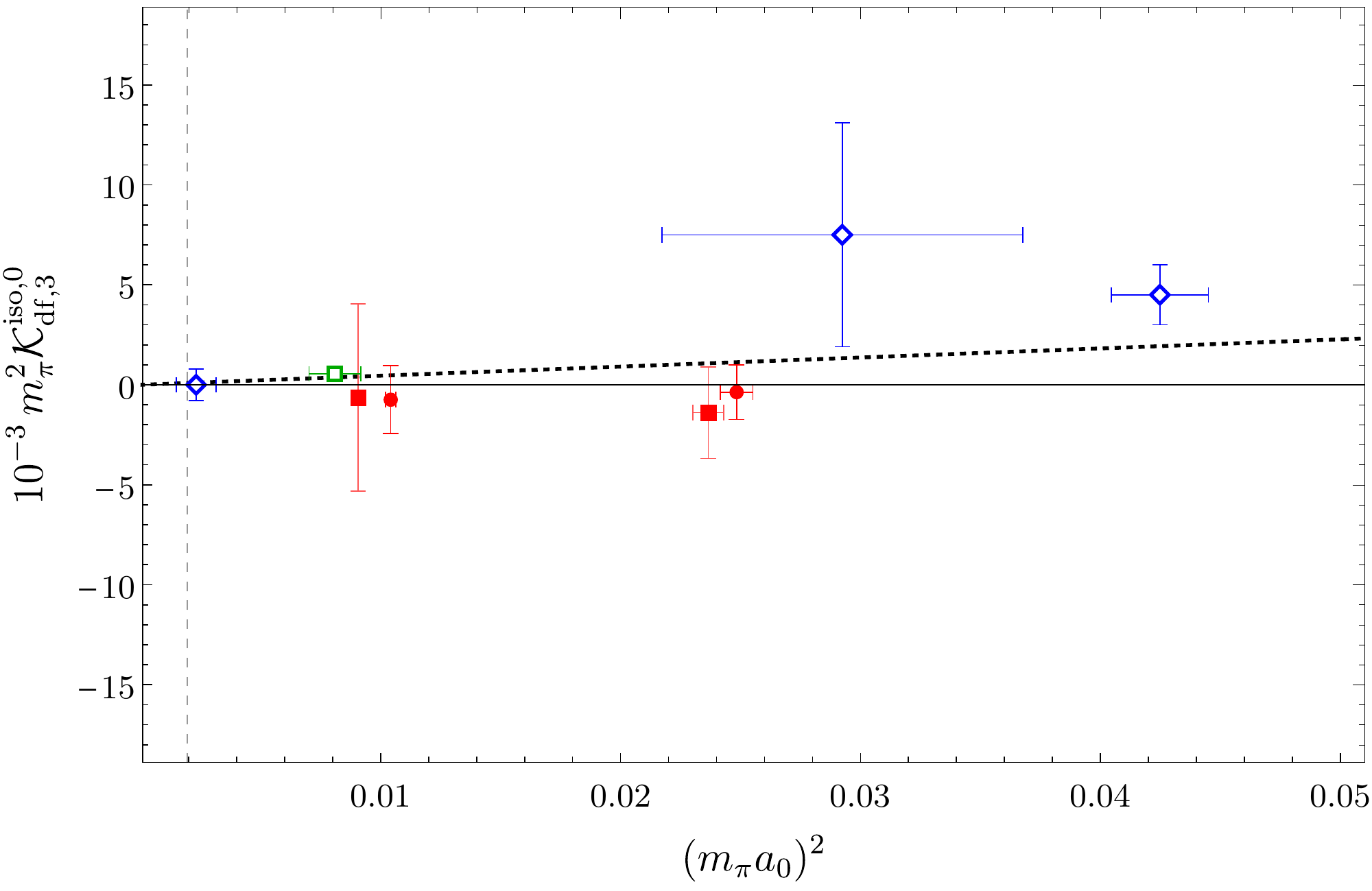}
    \includegraphics[width=0.495\textwidth,trim=0 0 0cm 0,clip]{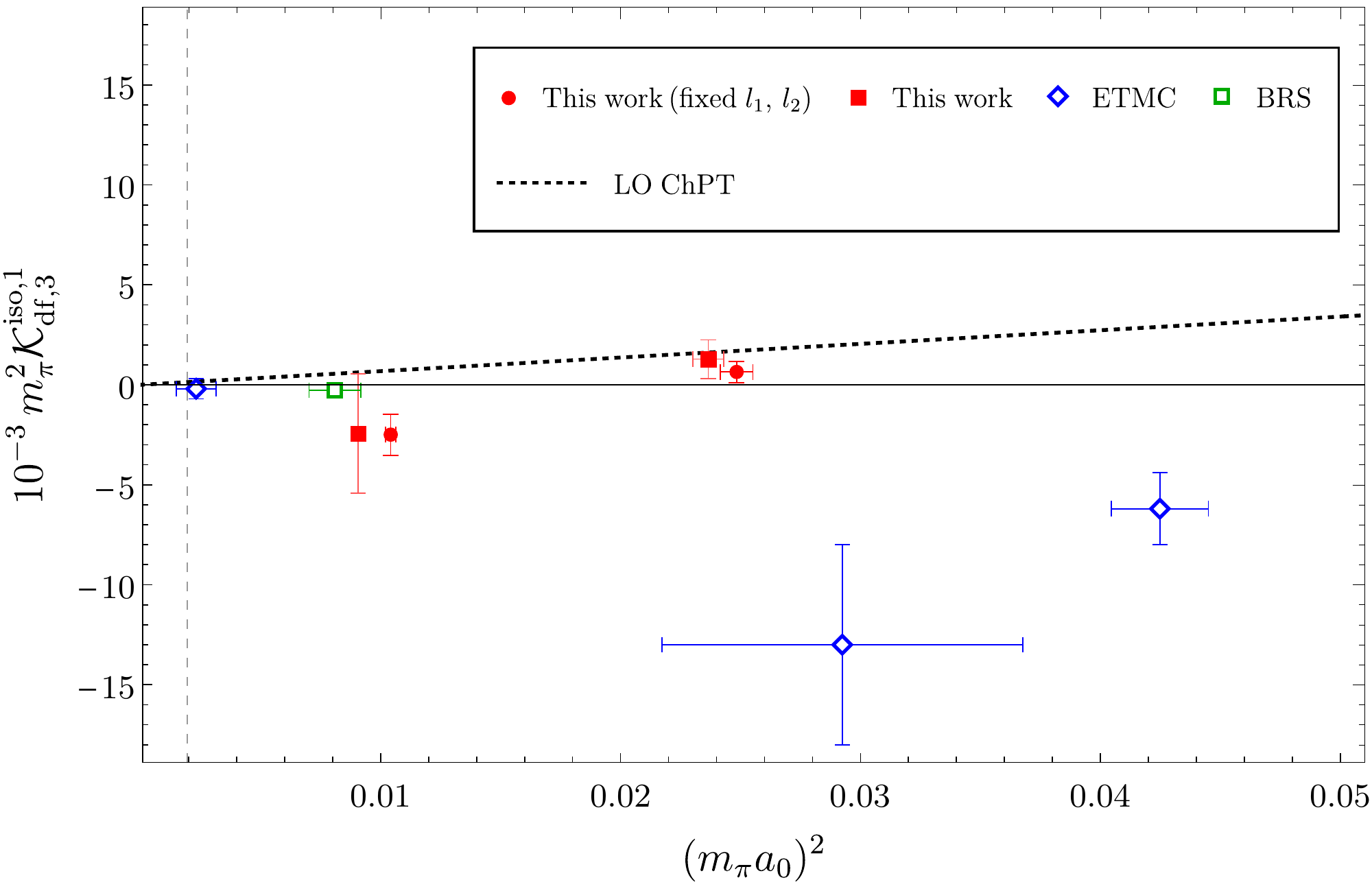}
    \caption{
    Three-body force versus the $I=2$ scattering length. Results from this work for fixed and varied two-body input is shown by filled red circles and squares, respectively. Expectations from leading order ChPT are shown by the gray line and those of earlier (RFT) determinations in blue~(ETMC~\cite{Fischer:2020jzp}) and green~(BRS~\cite{Blanton:2019vdk}) symbols. The red circles are slightly offset in the horizontal direction for legibility. The dashed vertical line shows the physical point.
    }
    \label{fig:T3barvsK3df}
\end{figure*}

An interesting question is whether we can extract the LECs parametrizing the two-body interactions directly from three-body energy levels. The three-body energy data set does not provide enough constraints to pin down both the LECs and the three-body terms. We are however able to fit the LECs when the three-body terms are set to zero. Setting $\bar{t}_{0,1}$ to zero makes sense, since their effect is rather small.  To stabilize the root finding routines used to predict the three-body energy levels as a function of the LECs, we constrained the region scanned for the LECs to a reasonable window, within one order of magnitude of the values determined from the two-body fits.
Procedurally this was accomplished
using a set of {\em relaxed} priors. We used a correlation matrix $\Sigma_\text{relaxed}=30^2\times\Sigma$, so that the
equivalent error bands on the LECs were at the level of 100\%, in effect constraining only the order of magnitude of the LECs.

The results for these fits are included in the last rows of Table~\ref{tab:fitsT3lecs}. We find that the values of the LECs are close to the ones generated from the two-body fits, albeit with larger error bars. This provides a good cross-check for the formalism and suggests that with enough three-body energy levels, we should be able to also constrain the two-body amplitudes.

To put the results on the three-body force in perspective, we compare our determination of the three-body term with those obtained in the literature~\cite{Blanton:2019vdk,Fischer:2020jzp} in Fig.~\ref{fig:T3barvsK3df}. In doing so, the matching of corresponding three-body terms can be made on the level of scattering amplitudes applying the procedure discussed in Sec.~\ref{subsec:qc-3}. We note that this yields an approximate identification ${\cal K}_{{\rm df},3}^{{\rm iso},0}\simeq6(\bar{t}_0+9\bar{t}_1)$ and ${\cal K}_{{\rm df},3}^{{\rm iso},1}\simeq54\bar{t}_1$.
We see reasonable agreement between different collaborations, not too different from the leading order ChPT prediction. This indicates the rapid progress made in the community in mapping out the three-body force.

\section{Conclusions}\label{sec:conclusions}

The field of three-body physics is rapidly advancing, fueled by progress on two fronts. On the one hand, precise energy levels are being produced in LQCD for interacting systems such as three pions or kaons.
On the other hand, formalisms that connect the finite-volume QCD spectrum and infinite-volume three-body scattering amplitude, called quantization conditions, are reaching maturity. Such progress has allowed the possibility of extracting quantitative information on the three-body force from first principles. 

In this work we apply the FVU formalism to analyze the spectrum obtained previously in Ref.~\cite{Culver:2019vvu}. We used a minimal parametrization for the three-body contact term and constrain the parameters from fits to the spectrum extracted using lattice QCD. We find that the heavy quark mass results are compatible with expectations from leading order ChPT, but our lower mass results are in tension with the predictions. Note that this is similar to other LQCD determinations of this term in the RFT framework~\cite{Blanton:2019vdk,Fischer:2020jzp}. The effects of the three-body force terms are small, in broad agreement with other lattice QCD extractions.
We also perform a fit to the three-body spectrum to constrain the parameters of the two-body amplitude. We find that the results are in agreement with the values extracted from the two-body spectrum, indicating that the two-body amplitude can also be determined consistently from the three-body data. While this is expected theoretically, it is an important feasibility check.

This study serves as a benchmark for the fitting strategy and the tools developed to generate predictions for the finite-volume three-body spectrum at maximal isospin. In this channel the effects of the three-body force are small, so to constrain it better we need more energy levels and/or better precision for the lattice data. This is likely to be done in the near future. Another direction that is being explored is to study other three-body channels where resonant amplitudes contribute~\cite{Hansen:2020zhy}. To this end, both lattice QCD data needs to be generated and the FVU quantization condition must be extended.

\bigskip
\acknowledgments
This material is based upon work supported by the National Science Foundation under Grant No. PHY-2012289 and by the U.S. Department of Energy under Award Number DE-SC0016582 (MD and MM) and DE-FG02-95ER40907 (AA, FXL, RB, CC). 
RB is also supported in part by the U.S. Department of Energy and ASCR, via a Jefferson Lab subcontract No. JSA-20-C0031.
CC is supported by UK Research and Innovation grant {MR/S015418/1}. The authors thank Fernando Romero-L\'opez and Peter Bruns for useful discussions.
\bigskip
\bibliographystyle{JHEP}
\bibliography{3pifitting}

\clearpage
\appendix

\section{Two-body summations}\label{sec:two-body-sums}

To set a baseline, the previous approach to two-body summations, as used in Refs.~\cite{Mai:2018djl, Mai:2019fba, Culver:2019vvu}, is reviewed. In Sec.~\ref{app:mIAM} the matching to ChPT is discussed motivating a new implementation for evaluating the two-body parameterizations  in Sec.~\ref{appsec:newtwo}.


\subsection{Previous two-body summation}
\label{sec:before}
In the following, the subscript ``$\infty$'' is used to distinguish infinite-volume quantities from their finite-volume counterparts; as outlined in the main text, starred {($^*$)} three-momenta and energies are defined in the two-body rest frame; likewise, tilde {(\,$\tilde{}$\,)} indicates the lattice rest frame, and momenta/energies without superscript are defined in the three-body rest frame.
Furthermore, $p,q,l$ label incoming, outgoing, and intermediate spectator momenta, respectively; $\tilde{P}$ is the momentum of the three-body system and $k^*$ labels the momentum of the pions from isobar decay, that are, of course, back-to-back, i.e., $(E_{\kvec^*},\kvec^*)$ and $(E_{\kvec^*},-\kvec^*)$. The zero-components of all momenta are always taken on-shell, i.e., $E_{\lvec}:= l^0=\sqrt{m_\pi^2+\lvec^2}$~\cite{Mai:2017vot}.

The infinite-volume propagator $\tau_\infty$ in Eq.~(4) of Ref.~\cite{Mai:2017bge}, adapted to current notation, reads
\begin{align}
\frac{\tau_\infty^{-1}}{\lambda^2}&=\frac{\sigma-M_0^2}{\lambda^2}-\Sigma_\infty \ ,\nonumber \\
\Sigma_\infty&=\int\frac{{\rm d}^3k^*}{(2\pi)^3}\,\frac{1}{2E_{\kvec^*}\,(\sigma-\sigma'+i\epsilon)} \ ,
\label{eq:tauinf}
\end{align}
where $\sigma':=4 E_{\kvec^*}^2$
and $\sigma=s+m_\pi^2-2\sqrt{s}\,E_{\lvec}$ is the invariant two-body sub-energy squared (in the main text, $\sigma$ is called $\sigma_{\pvec}$ owing to the fact that $\pvec $ and $\qvec $ are used to name spectator momenta). We refer to the term $\Sigma$ as ``self energy'' in the following although this formalism is a-priori not based on any diagrammatic expansion. Furthermore, $M_0$ and $\lambda$ in Eq.~\eqref{eq:tauinf} are fit parameters that may be adjusted to two-body input; indeed, the two-to-two scattering $T$-matrix is simply 
\begin{align}
T_{22, \infty}=vSv=-v\frac{1}{D}v
=-\lambda\tau_\infty\lambda
\label{eq:t22}
\end{align}
with $S$ and $D$ from Ref.~\cite{Mai:2017vot}. We also indicate the restriction of the general vertex $v\equiv v(q_i,q_j)$, that depends on the four-momenta of the decay pions $q_i,\,q_j$~\cite{Mai:2017vot}, to the $S$-wave case $\lambda\equiv \lambda(\sigma)$ that here depends only on $\sigma$. The vertex $v$ depends on invariants formed by the three available (isobar and decay) four-momenta. This ensures three-body unitarity is maintained when $v$ is evaluated in different frames (three-body and two-body rest frames)~\cite{Sadasivan:2020syi}.In previous studies~\cite{Mai:2018djl, Mai:2019fba, Culver:2019vvu} a form factor for regularization was included in the definition of $v$ which we can drop in the current formulation as explained in Sec.~\ref{appsec:newtwo}; for the time being we assume that the integral in Eq.~\eqref{eq:tauinf} is simply regularized by a cutoff  (in Sec.~\ref{appsec:newtwo} we construct convergent expressions). 

The finite-volume version of the isobar-spectator propagator is obtained from Eq.~\eqref{eq:tauinf} by imposing periodic boundary conditions in the lattice rest frame leading to discretized momenta. For elongations $\eta$ in $z$-directions, i.e., $L_x=L_y=L$, $L_z=\eta L$ this implies a discrete set of allowed three-momenta $\tkvec\in \mathcal{S}_{L\eta}:=\{2\pi/L(n_1,n_2,n_3/\eta)|(n_1,n_2,n_3)\in \mathds{Z}^3\}$. The finite-volume propagator is diagonal in spectator momentum and reads
\begin{align}
\frac{\tau^{-1}_{\tPvec L\eta}}{\lambda^2}&=\frac{\sigma-M_0^2}{\lambda^2}-\Sigma_{\tPvec L\eta}
\nonumber \\
\Sigma_{\tPvec L\eta}&=
\frac{1}{\eta L^3} 
\sum_{\kvec^*}
\frac{\tilde J\,J\,}{2E_{\kvec^*}}
\frac{1}{\sigma-\sigma'}
\label{eq:DSE}
\end{align}
where $\kvec^{*}\equiv \kvec^{*}(\kvec(\tkvec),\lvec(\tlvec))$,
with the dependence on spectator momentum $\lvec$ explicitly denoted. 
The boost momenta are $\tPvec$ from lattice to three-body rest frame, leading to the Jacobian $\tilde J$, and $-\lvec$ from three-body rest frame to two-body rest frame, leading to the Jacobian $J$ in Eq.~\eqref{eq:DSE}.

To further discuss the kinematics, we consider the incoming and outgoing pion momenta $\tilde \qvec_j$ and $\tilde \pvec_j$, $j=1,2,3$. The $3\pi^+$ system has then momentum $\tPvec=\tqvec_1+\tqvec_2+\tqvec_3=
\tpvec_1+\tpvec_2+\tpvec_3$. Momenta in the three-body rest frame are~\cite{Doring:2012eu}
\begin{align}
\lvec = \tlvec+\left[
\left(\frac{E_{\tPvec}}{\sqrt{s}}-1\right)\frac{\tlvec\cdot\tPvec}{\tPvec^2}-\frac{E_{\tlvec}}{\sqrt{s}}\right]\tPvec\,,
\label{eq:moveframe}
\end{align}
and analogously for the other momenta $\kvec$, $\pvec$ and $\qvec$.
In Eq.~\eqref{eq:moveframe}, $E_{\tPvec}=\sqrt{s+\tPvec^2}$. The boost of Eq.~\eqref{eq:moveframe} leads to the Jacobian
\begin{align}
\tilde J=
\left|\frac{d\lvec}{d\tlvec}\right|
=
\frac{E_{\tPvec}}{\sqrt{s}}-\frac{\tlvec\cdot\tPvec}{\sqrt{s}\,E_{\tlvec}} \ .
\label{eq:JacobianTilde}
\end{align}

The isobar is not at rest in the three-body rest frame. Thus, an additional boost (by $-\lvec$) has to be performed for the pertinent summation of momenta $\kvec^*$ in the self energy of the isobar of Eq.~\eqref{eq:DSE}. This is detailed in Eqs.~(11, 12) of Ref.~\cite{Mai:2017bge} and reads in the current notation
\begin{align}\label{eq:boost}
\kvec^*
=
\kvec+
\lvec\left[\frac{\kvec\cdot \lvec}{\lvec^2}\left(\frac{\sqrt{\sigma}}{\sqrt{s}-E_{\lvec}}-1\right)+
\frac{\sqrt{\sigma}}{2(\sqrt{s}-E_{\lvec})}\right] \ ,
\end{align}
leading to the Jacobian
\begin{align}\label{eq:Jacobian}
J=\frac{\sqrt{\sigma}}{\sqrt{s}-E_{\lvec}} \ .
\end{align}


\subsection{Matching to ChPT}
\label{app:mIAM}
We use the Inverse Amplitude Method (IAM)~\cite{Truong:1988zp, Dobado:1996ps} for the isospin $I=2$ $\pi\pi$ scattering amplitude, with a modification for improved sub-threshold behavior (mIAM)~\cite{GomezNicola:2007qj}. For this subsection, we denote the standard Mandelstam variables by $\sigma$, $t$, $u$ instead of $s$, $t$, $u$ to avoid notation clash with other parts of the paper. The starting point is given by the ChPT result at leading (LO) and next-to-leading order (NLO),
\begin{align}
T^I_2(\sigma, t,u) \quad\text{and}\quad T^I_4(\sigma, t,u) \ ,
\label{app:explit}
\end{align}
for $I=2$ scattering. From now on we drop the index $I$ for readability. 
The explicit expressions are given in Ref.~\cite{Gasser:1983yg}.
In $S$-wave ($\ell=0$), the unitary amplitude $T_{\rm mIAM}^{\ell}$ is written as
\begin{align}
T^{0}_{\rm mIAM}=\frac{(T^{0}_2)^2}{T^{0}_2-T^{0}_4+A}\,,
\label{eq:tmIAM}
\end{align}
where $A$ is constructed such that the Adler zero is at its NLO position, i.e.,
\begin{align}
\label{app:modi}
    A(\sigma)=T_4^0(\sigma_2&)-\\
    &\frac{(\sigma_2-\sigma_A)(\sigma-\sigma_2)}{\sigma-\sigma_A}\left(T_2^{0\prime}(\sigma_2)-T_4^{0\prime}(\sigma_2)\right)\,,
    \nonumber
\end{align}
where $\sigma_2$ and $\sigma_A$ are zeroes of $T_2^0$ and $T_2^0-T_4^0$, respectively, see Ref.~\cite{GomezNicola:2007qj} for more details.

In Eqs.~\eqref{eq:tmIAM} and \eqref{app:modi}, $T_2^{0}$ ($T_4^{0}$) is the LO (NLO) partial-wave contribution, obtained from the corresponding plane-wave expressions of Eq.~\eqref{app:explit} according to
\begin{align}
T_n^{0}(\sigma)=\frac{1}{32N\pi}\int\limits_{-1}^1\mathrm{d}x\,P_0\,T_n(\sigma,t,u) \,,
\label{eq:pwa}
\end{align}
for $n=2,4$ and $x=\cos\theta$ where $\theta$ is the scattering angle. There is also a symmetry factor of $N=2$ for identical particles, and $P_0=1$ is the $S$-wave Legendre polynomial. In particular, $T_2=32\pi T^{0}_2$ as the LO contribution is angle independent.

In a next step, we match this to a K-matrix formalism which is easier to implement into the three-body framework. For this we note that the NLO contribution to the $\pi\pi$ amplitude can be split as
\begin{align}
T_4(\sigma,t, u)=\bar T_4(\sigma,t, u)+\frac{1}{2}\,(T_2(\sigma))^2\bar J_{\pi\pi}(\sigma)
\end{align}
or
\begin{align}
T_4^{0}(\sigma)=\bar T_4^{0}(\sigma)+16\pi\, (T_2^{0}(\sigma))^2\,\bar J_{\pi\pi}(\sigma) \,,
\label{eq:t4sep}
\end{align}
where $\bar T_4$ and $\bar T_4^{0}$ are real for $s>4m_\pi^2$ and only the $\pi\pi$ loop $\bar J_{\pi\pi}$ provides an imaginary part in the physical region. It reads~\cite{GomezNicola:2001as} 
\begin{align}
\bar J_{\pi\pi}=\frac{1}{16\pi^2}\left[2+\hat\sigma\log\frac{\hat\sigma-1}{\hat\sigma+1}\right] \,,
\label{eq:Jbar}
\end{align}
where
\begin{align}
\hat\sigma=\frac{2k_\text{cm}}{\sqrt{\sigma}} \,,\quad k_\text{cm}=\sqrt{\frac{\sigma}{4}-m_\pi^2} \,.
\end{align}
We can now determine the (real-valued) $K$-matrix~\footnote{Strictly speaking, for a $K$-matrix formalism $\text{Re } \bar J_{\pi\pi}$ should be absorbed in $K$, as well; somewhat sloppily, we still refer to $K$ as ``$K$-matrix''.} by equating 
\begin{align}
T^{0}_{\rm mIAM}=(K^{-1}-\rho)^{-1}
\label{eq:tK}
\end{align}
with Eq.~\eqref{eq:tmIAM}. Using Eqs.~\eqref{eq:t4sep} this leads to 
\begin{align}
K^{-1}&=\frac{T^0_2-\bar T^0_4+A}{(T^0_2)^2} \,,\quad
\rho=16\pi \bar J_{\pi\pi} \,,
\label{app:KMatrix}
\end{align}
which, of course, only depends on $\sigma$.
We obtain the plain-wave amplitude $T_\text{mIAM}$ from Eqs.~\eqref{eq:tK}, \eqref{eq:pwa} and can set it equal to the two-to-two scattering matrix of Eq.~\eqref{eq:t22},
\begin{align}
-
\left(\lambda\tau_\infty\lambda\right)^{-1}=
T^{-1}_{22,\infty}=T^{-1}_\text{mIAM}=
\frac{K^{-1}}{32\pi}-\frac{\bar J_{\pi\pi}}{2}\,.
\label{app:equal}
\end{align}
Indeed, the conventions of both $T$-matrices are identical as an explicit evaluation of imaginary parts shows:
\begin{align}
&\text{Im }T^{-1}_{22,\infty}~=\text{Im }\Sigma_\infty
=-\frac{k_{\rm cm}}{16\pi\sqrt{\sigma}}\,,
\nonumber \\
&\text{Im }T^{-1}_{\text{mIAM}}=-\frac{16\pi}{32\pi}\,\text{Im }\bar J_{\pi\pi}=-\frac{k_{\rm cm}}{16\pi\sqrt{\sigma}} \,.
\label{eq:problem}
\end{align}
For the quantization condition, we also need the equivalent of the $K$-matrix of Eq.~\eqref{app:KMatrix} in the plane-wave basis. This explains the additional factor of $32\pi$ in Eq.~\eqref{eq:qc} of the main text using Eq.~\eqref{eq:pwa}.

A final remark on symmetry factors is in order because we match our dispersive three-body framework to the Feynman-diagrammatic mIAM approach, and keeping track of these factors is important. The symmetry normalization of the two-body amplitude happens in the partial-wave decomposition Eq.\eqref{eq:pwa} ($N=2$) such that the partial-wave amplitudes are connected to physical observables in the standard way; for example, the $S$-wave scattering length is $a_0 m_\pi=T^0_{\rm mIAM}(\sigma=4m_\pi^2)$ and the phase shift reads $\tan\delta=\frac{\text{Im }T^0_{\rm mIAM}}{\text{Re }T^0_{\rm mIAM}}$.
When returning from the partial-wave $T^0_\text{IAM}$ to the plane-wave $T_\text{IAM}$ in Eq.~\eqref{app:equal} the corresponding factor $32\pi$ contains the $N=2$ of Eq.~\eqref{eq:pwa}, i.e., $T_\text{IAM}$ is \emph{not} symmetry normalized. The same is true for $T_{22,\infty}$: In the notation of Ref.~\cite{Mai:2017vot}, the matrix element for incoming pions at momenta $p_1,p_2$ and outgoing momenta $q_1,q_2$ is given as 
\begin{align}
    \bra{q_1,q_2}{\cal T}\ket{p_1,p_2}
    &=\frac{1}{2!}v(q_1,q_2)S(\sigma)v(p_1,p_2) \\
    &=-\frac{1}{2}\lambda(\sigma)\tau_\infty(\sigma)\lambda(\sigma) =\frac{1}{2} T_{22,\infty}
    \nonumber \,,
\end{align}
i.e. the symmetry normalization is \emph{not} contained in the definition of $T_{22,\infty}$. Furthermore, note that $v=2\tilde v$ where the factor of 2 accounts for the possibilities to connect a decay vertex $\tilde v$ to two external pions~\cite{Mai:2017vot}. In other words, $v$ contains the same multiplicity as a Feynman vertex would carry from the two possible contractions with external fields.


\subsection{Regularization-free, accelerated two-body summation}
\label{appsec:newtwo}
It is easy to see that Eq.~\eqref{eq:DSE} diverges logarithmically. In the past~\cite{Mai:2018djl, Mai:2019fba, Culver:2019vvu} the same  cutoff $\Lambda$ was chosen for both integration and summation in Eqs.~\eqref{eq:tauinf} and \eqref{eq:DSE} such that the real parts approximately match (indicated with ``!''), 
\begin{align}
\text{Re }\Sigma_\infty(\sigma)\,\stackrel{!}{\approx}-\frac{1}{2}\text{Re }\bar J_{\pi\pi}(\sigma)-\frac{d}{4\pi^2}
\label{eq:oldmatch}
\end{align}
which holds well over a large range of $\sigma$ for the choice $d=0.86$ and $\Lambda=42m_\pi$~\cite{Mai:2018djl}.  Equation~\eqref{eq:oldmatch} completed the matching procedure. 

However, one can take a step back and question the form of the self energy in Eq.~\eqref{eq:tauinf}. After all, its specific form was chosen to be able to match to time-ordered perturbation theory with explicit isobar fields~\cite{Mai:2017vot}, which is not our goal here. 
In Ref.~\cite{Mai:2017vot}, different forms of the isobar-spectator are derived from its imaginary part given in Eq.~\eqref{eq:problem}. In particular, $n$-times subtracted dispersion relations provide factors of $(\sigma/\sigma')^n$ in the integrand for the self energy such that already with $n=1$ one obtains a convergent expression in the present case. As convergence is drastically improved, no matching with hard cutoffs is needed any more and the numerical summation can be cut at much lower values, greatly improving the speed of fitting that was previously limited by the two-body summation. With the improvements discussed in the following, typical speedups of a factor of 10 are achieved.

Consider the twice-subtracted dispersion relation with subtraction point arbitrarily chosen at $\sigma_0=0$,
\begin{align}
\Sigma_\infty^{(2)}(\sigma)
& =\Sigma_\infty^{(2)}(0)
+\sigma\,\Sigma_\infty^{(2)\prime}(0)
+\frac{\sigma^2}{\pi}\int\limits_{4m_\pi^2}^\infty \mathrm{d}\bar\sigma\,
\frac{\text{Im }\Sigma_\infty(\bar\sigma)}{\bar\sigma^2\left(\bar\sigma-\sigma-i\epsilon\right)}\nonumber \\
& =
-\frac{1}{2}\frac{\sigma}{96\,m_\pi^2\pi^2}
+\frac{\sigma^2}{64\pi^2}\int\limits_0^\infty
\mathrm{d}k^*\frac{k^{*2}}{E_{\kvec^*}^5}\,\frac{1}{\sigma-\sigma'+i\epsilon} \ ,
\label{eq:twice}
\end{align}
where $\sigma'=4E_{\kvec^*}^2$ and the second equal sign is obtained with the imaginary part from Eq.~\eqref{eq:problem}, a variable substitution, and by matching $\Sigma_\infty^{(2)}$ and its derivative at $\sigma=0$ to $-1/2\,\bar J_{\pi\pi}$ from Eq.~\eqref{eq:Jbar}. It can be easily shown that these two functions are identical for \emph{all} $\sigma$,
\begin{align}
\Sigma_\infty^{(2)}(\sigma)=-\frac{1}{2}\,\bar J_{\pi\pi}(\sigma) 
\approx
\Sigma_\infty(\sigma)+\frac{d}{4\pi^2}\ ,
\end{align}
where we have, for completeness, also quoted the relation to the original expression according to Eq.~\eqref{eq:oldmatch}. The advantage of using the dispersion relation for the matching to $\bar J_{\pi\pi}$ is now apparent: the matching is exact, not approximate, it is independent of matching parameters (cut-off $\Lambda$ and $d$), and the corresponding integration converges quickly, as Eq.~\eqref{eq:twice} shows. 

By writing 
\begin{align}
\Sigma_\infty^{(2)}
=-\frac{\sigma}{192m_\pi^2\pi^2}+\int\frac{\mathrm{d}^3 k^*}{(2\pi)^3}\,\frac{1}{2E_{\kvec^*}}\left(\frac{\sigma}{\sigma'}\right)^2\frac{1}{\sigma-\sigma'+i\epsilon} \ ,
\end{align}
the formal similarity with Eq.~\eqref{eq:tauinf}, up to the factor $(\sigma/\sigma')^2$, becomes apparent. By imposing periodic boundary conditions as in Sec.~\ref{sec:before}, the finite-volume self energy becomes
\begin{align}
\Sigma^{(2)}_{\tPvec L\eta}&=
-\frac{\sigma}{192m_\pi^2\pi^2}+
\frac{1}{\eta L^3} 
\sum_{\kvec^*}
\frac{\tilde J\,J\,}{2E_{\kvec^*}}
\left(\frac{\sigma}{\sigma'}\right)^2
\frac{1}{\sigma-\sigma'} \ .
\label{eq:twicefinite}
\end{align}
This expression may be used instead of Eq.~\eqref{eq:DSE} as it converges rapidly and provides exact matching to the ChPT loop $\bar J_{\pi\pi}$.

The difference between $\Sigma_{\tPvec L\eta}^{(2)}$ and $\Sigma_{\tPvec L\eta}$ from Eq.~\eqref{eq:DSE} is exponentially suppressed for all $\sigma$. Indeed, whenever a finite-volume pole arises at $\sigma-\sigma'=0$ (remember that $\sigma'$ is discrete in finite volume), we have $(\sigma/\sigma')^2=1$, i.e., the finite-volume poles in both expressions have the same residues. 
Furthermore, the difference is of the form 
\begin{align}
\Sigma_{\tPvec L\eta}-\Sigma_{\tPvec L\eta}^{(2)}=A_0+A_1\sigma
\label{eq:AB}
\end{align}
as an inspection of the arguments of the summations shows:
\begin{align}
\frac{1}{\sigma-\sigma'}-\left(\frac{\sigma}{\sigma'}\right)^2 \frac{1}{\sigma-\sigma'}=-\frac{\sigma}{\sigma^{\prime 2}}-\frac{1}{\sigma'} \ .
\end{align}
The exponentially suppressed difference can be sizable and becomes, in general, larger for more subtractions in the dispersion relation; if one aims at regaining a particular form of the self energy, one can reduce the difference. To do this we solve explicitly for $A_1$, with the crucial advantage that it has to be calculated only \emph{once} for given $L,\eta$, three-body boost $\tPvec$, and two-body boost by $-\lvec$. The quantity $A_1$ can then be recycled, e.g., in fitting, where the entire $\sigma$-dependence of $\Sigma$ needs to be known. In other words, one can still take advantage of the speedup provided by $\Sigma_{\tPvec L\eta}^{(2)}$ at the cost of having to determine $A_1$, or $A_0$ and $A_1$, once for every $L$, $\eta$, $\tPvec$, $-\lvec$.
The quantity $A_1$ is easiest determined in the limit $\sigma\to-\infty$. Using Eqs.~\eqref{eq:twice} and \eqref{eq:twicefinite},
\begin{align}
A_1=\frac{1}{4\pi^2}\int\limits_0^\infty \mathrm{d}k^*\,\frac{k^{*2}}{\sigma^{\prime2}E_{\kvec^*}}-\frac{1}{2\eta L^3}\sum_{\kvec^*}\frac{\tilde J J}{\sigma^{\prime 2}E_{\kvec^*}} \,,
\end{align}
which is indeed exponentially suppressed.

This concludes the derivation. Still, one can write the self energy, modified by the $A_1$-term, in the surprisingly simple form
\begin{align}
\Sigma_{\tPvec L\eta}^{(1)}:=\Sigma_{\tPvec L\eta}^{(2)}+A_1\sigma
=\frac{1}{\eta L^3} 
\sum_{\kvec^*}
\frac{\tilde J\,J\,}{2E_{\kvec^*}}
\frac{\sigma}{\sigma'}
\frac{1}{\sigma-\sigma'} \,.
\label{eq:onesubtract}
\end{align}
The factor $\sigma/\sigma'$ appears here linearly. Indeed, a closer inspection shows that we could have obtained Eq.~\eqref{eq:onesubtract} directly by imposing periodic boundary conditions on a once-subtracted dispersion relation, i.e., by starting the derivation with 
\begin{align}
\Sigma_\infty^{(1)}(\sigma) =\Sigma_\infty^{(1)}(0)
+\frac{\sigma}{\pi}\int\limits_{4m_\pi^2}^\infty \mathrm{d}\bar\sigma\,
\frac{\text{Im }\Sigma_\infty(\bar\sigma)}{\bar\sigma\left(\bar\sigma-\sigma-i\epsilon\right)}
\label{app:onceinf}
\end{align}
instead of Eq.~\eqref{eq:twice}. Of course, by using directly Eq.~\eqref{eq:onesubtract} one partially looses the speedup of the twice-subtracted results. But, at least, Eq.~\eqref{eq:onesubtract} is still convergent in contrast to Eq.~\eqref{eq:DSE}. For the purpose of this study, we find the speed-up provided by Eq.~\eqref{eq:onesubtract} sufficient and trade the slight loss of speed for not having to calculate $A_1$ separately for each $L$, $\eta$, $\tPvec$, $-\lvec$. The exponentially suppressed term $A_0$ of Eq.~\eqref{eq:AB} is tiny and leads to changes of the three-body finite-volume energies in the sixth significant digit.

\end{document}